\RequirePackage{fix-cm}
\documentclass[11pt,a4paper]{article}
\pdfoutput=1
\RequirePackage{jheppub}

\RequirePackage{atlasphysics}
\RequirePackage{lineno}
\RequirePackage{booktabs}
\RequirePackage{multirow}
\RequirePackage[section]{placeins}
\RequirePackage{verbatim}
\RequirePackage{morefloats}
\RequirePackage{float}
\RequirePackage[labelfont=bf,font=small,labelsep=period]{caption}
\renewcommand{\figurename}{Figure}
\RequirePackage{preprintcover}
\PreprintCoverPaperTitle{Measurement of the top quark pair production charge asymmetry in proton--proton collisions at $\sqrt{s}=7$~TeV using the ATLAS detector}  % paper title
\PreprintIdNumber{CERN-PH-EP-2013-177}  % CERN preprint number
\PreprintCoverAbstract{This paper presents a measurement of the top quark pair (\ttbar) production charge asymmetry \Ac\ using \lumi\ of proton--proton collisions at a centre--of--mass energy 
 $\sqrt{s} =  7$~TeV collected by the ATLAS detector at the LHC.  A \ttbar\--enriched sample of events with a single lepton (electron or muon), missing transverse momentum and at 
 least four high  transverse momentum jets, of which at least one is tagged as coming from a $b$--quark, is selected. A likelihood fit is used to reconstruct the \ttbar~event kinematics.
 A Bayesian unfolding procedure is employed to estimate \Ac\ at the parton--level. The measured value of the \ttbar\ production charge asymmetry is \mbox{\resultincl}, where the
 uncertainty includes both the statistical and the systematic components.
 Differential \Ac\ measurements as a function of the invariant mass, the rapidity and the transverse momentum of the \ttbar--system are also presented. In addition, 
 \Ac\ is measured for a subset of events with large \ttbar\ velocity, where physics beyond the Standard Model could contribute. All measurements are consistent with the Standard Model
 predictions.}  % paper abstract
\PreprintJournalName{JHEP}  % journal name

\def\TeV{\ifmmode {\mathrm{\ Te\kern -0.1em V}}\else
                   \textrm{Te\kern -0.1em V}\fi}%
\def\GeV{\ifmmode {\mathrm{\ Ge\kern -0.1em V}}\else
                   \textrm{Ge\kern -0.1em V}\fi}%
\def\MeV{\ifmmode {\mathrm{\ Me\kern -0.1em V}}\else
                   \textrm{Me\kern -0.1em V}\fi}%
\def\keV{\ifmmode {\mathrm{\ ke\kern -0.1em V}}\else
                   \textrm{ke\kern -0.1em V}\fi}%
\def\eV{\ifmmode  {\mathrm{\ e\kern -0.1em V}}\else
                   \textrm{e\kern -0.1em V}\fi}%
\def\DR{\ensuremath{\Delta R \equiv \sqrt{(\Delta \eta)^2 + (\Delta \phi)^2}}}%
\def\ST{\ensuremath{S(\Truth{}) \equiv \abs{C(\Truth{})-C(\Truth{}^*)}}}%

\newcommand{\jimmy}{{\sc JIMMY}}
\newcommand{\herwig}{{\sc HERWIG}}
\newcommand{\powheg}{{\sc POWHEG}}
\newcommand{\pythia}{{\sc PYTHIA}}
\newcommand{\alpgen}{{\sc ALPGEN}}

\newcommand{\mcnlo}{{\sc MC@NLO}}

\newcommand{\acermc}{{\sc AcerMC}}
\newcommand{\mttbar}{\ensuremath{m_{t\bar{t}}}}
\newcommand{\yttbar}{\ensuremath{|y_{t\bar{t}}|}}
\newcommand{\ptttbar}{\ensuremath{p_{\mathrm{T},t\bar{t}}}}
\newcommand{\lumi}{4.7~\ifb}
\newcommand{\Yt}{\ensuremath{|y_t|}}
\newcommand{\Ytbar}{\ensuremath{|y_{\tbar}|}}
\newcommand{\DY}{\ensuremath{\Delta |y|}\ }

\newcommand{\Ac}{\ensuremath{A_{\rm{C}}}}

\newcommand{\resultincl}{\ensuremath{\Ac = 0.006 \pm 0.010 }}
\newcommand{\resultinclbeta}{\ensuremath{\Ac = 0.011 \pm 0.018 }}

\providecommand{\abs}[1]{\lvert#1\rvert}

\newcommand{\Truth}{\ensuremath{\mathbf{T}}}
\newcommand{\Data}{\ensuremath{\mathbf{D}}}
\newcommand{\TrasfMatrix}{\ensuremath{\mathcal{M}}}
\newcommand{\Real}{\ensuremath{\mathbb{R}}}
\newcommand{\Integer}{\ensuremath{\mathbb{N}}}
\newcommand{\conditionalProb}[2]{\ensuremath{p\left(#1|#2\right)}}
\newcommand{\conditionalLhood}[2]{\ensuremath{\mathcal{L}\left(#1|#2\right)}}
\newcommand{\prior}{\ensuremath{\pi{}\left(\Truth{}\right)}}

\title{Measurement of the top quark pair production charge asymmetry in proton--proton collisions at $\sqrt{s}=7$~TeV using the ATLAS detector}
\author[a]{The ATLAS Collaboration}
\affiliation[a]{CERN, 1211 Geneva 23, Switzerland}
\emailAdd{atlas.publications@cern.ch}

\abstract{This paper presents a measurement of the top quark pair (\ttbar) production charge asymmetry \Ac\ using \lumi\ of proton--proton collisions at a centre--of--mass energy 
 $\sqrt{s} =  7$~TeV collected by the ATLAS detector at the LHC.  A \ttbar\--enriched sample of events with a single lepton (electron or muon), missing transverse momentum and at 
 least four high  transverse momentum jets, of which at least one is tagged as coming from a $b$--quark, is selected. A likelihood fit is used to reconstruct the \ttbar~event kinematics.
 A Bayesian unfolding procedure is employed to estimate \Ac\ at the parton--level. The measured value of the \ttbar\ production charge asymmetry is \mbox{\resultincl}, where the
 uncertainty includes both the statistical and the systematic components.
 Differential \Ac\ measurements as a function of the invariant mass, the rapidity and the transverse momentum of the \ttbar--system are also presented. In addition, 
 \Ac\ is measured for a subset of events with large \ttbar\ velocity, where physics beyond the Standard Model could contribute. All measurements are consistent with the Standard Model
 predictions.}

\arxivnumber{hep-ex:1311.6724}
\notoc
\begin{document}
\maketitle
\flushbottom
%\begin{frontmatter}
%\linenumbers

\section{Introduction}
\label{Introduction}
The measurement of the \ttbar\ production charge asymmetry represents an important test of quantum 
chromodynamics (QCD)
at high energies and is also an ideal place to observe effects of
possible new physics processes beyond the Standard Model (BSM). 
Several BSM processes can alter this asymmetry
~\cite{Jung:2011zv, AXI,Djouadi:2009nb, KK, Jung:2009jz, Shu:2009xf, JA:2011, AguilarSaavedra:2011hz, Dorsner:2009mq, Grinstein:2011yv, Ligeti:2011vt, Ferrario:2009bz, Frampton:2009rk}, 
either with anomalous vector or axial--vector couplings (i.e. axigluons) or via interference with the Standard Model (SM). 
Different models also predict different asymmetries as a function of the invariant mass \mttbar~\cite{AguilarSaavedra:2011ci}, the transverse momentum \ptttbar\ and the rapidity \yttbar\ of the
\ttbar--system.

At leading order (LO), \ttbar\ production at hadron colliders is predicted to be symmetric under the 
exchange of top quark and antiquark. At next--to--leading order (NLO), the process 
$q\bar{q}\to t\bar{t}g$ exhibits an asymmetry in the rapidity distributions of the top quark and antiquark, 
due to interference between initial-- and final-- state gluon emission. 
In addition, the $q\bar{q}\to t\bar{t}$ process itself possesses an asymmetry due to the interference 
between the Born and the NLO diagrams. The $qg$ production process is also asymmetric, but its contribution is much smaller than the \qqbar\ one.
The production of \ttbar\ events by gluon fusion, $gg \to t\bar{t}$, is symmetric.
At the Tevatron proton--antiproton collider, where \ttbar\ events are predominantly produced by \qqbar\ annihilation, 
top quarks are preferentially emitted in the direction of the incoming quark while the top antiquarks are emitted preferentially in the direction of the incoming 
antiquark~\cite{Kuhn:1998kw,Kuhn:1998jr,Bernreuther:2010ny,Ahrens:2011uf,Hollik:2011ps,Kuhn:2011ri,Bern:2012}.
The \ttbar\ asymmetry at the Tevatron is therefore measured as a forward--backward asymmetry,
\begin{linenomath}
\begin{equation}
A_{\rm{FB}}= \frac{N(\Delta y>0)-N(\Delta y<0)}{N(\Delta y>0)+N(\Delta y<0)},
\nonumber
\end{equation}
\end{linenomath}
where $\Delta y \equiv y_t-y_{\bar{t}}$ is the difference in rapidity between top quarks and antiquarks, and $N$ represents the number of events with $\Delta y$ being positive or negative. 
The interest in this measurement has grown after CDF and D0 
collaborations reported $A_{\rm{FB}}$ measurements significantly larger than the SM predictions, in both the inclusive and differential 
case as a function of \mttbar\ and \yttbar~\cite{CDF1,CDF2,CDF3,D01,D02}.

In proton--proton ($pp$) collisions at the LHC, the dominant mechanism for \ttbar\ production is the $gg$ 
fusion process, while production via \qqbar\ or $qg$ interactions is small. 
Since the colliding beams are symmetric, $A_{\rm{FB}}$ is no longer a useful observable.
However, \ttbar\ production via \qqbar\ or $qg$ processes is asymmetric under top quark--antiquark exchange, and, in addition, the valence quarks carry, on average, a larger momentum fraction than antiquarks from the sea.
Hence for \qqbar\ or $qg$ production processes at the LHC, QCD predicts a small excess of centrally produced top antiquarks while top quarks are produced, 
on average, at higher absolute rapidities.
Therefore, the \ttbar\ production charge asymmetry \Ac\ is defined as~\cite{Jung:2011zv,Diener:2009ee} 
\begin{linenomath}
\begin{equation}
\Ac = \frac{N(\Delta |y| >0) - N(\Delta |y| <0)}{N(\Delta |y| >0) + N(\Delta |y| <0)},
\label{Ac}
\end{equation}
\end{linenomath}
where $\DY \equiv \Yt - \Ytbar$  is the difference between the absolute value of the top quark rapidity \Yt\ and the absolute value of the top antiquark rapidity \Ytbar.

The SM prediction for the \ttbar\ production charge asymmetry at the LHC is $A_{\rm{C}}^{\text{SM}}=0.0123\pm0.0005$~\cite{Bern:2012}, computed at NLO in QCD including electroweak corrections.
Recent asymmetry measurements at the LHC~\cite{CMS_ljets,CMS_ljets2,ATLAS_ljets} did not report any significant deviation 
from the SM predictions in either the inclusive or differential \Ac\ measurements. Agreement with the SM \Ac\ predictions at the LHC is compatible with the larger than expected
$A_{\rm{FB}}$ values measured at the Tevatron for the most 
general new physics scenarios~\cite{AguilarSaavedra:2012prl}, but creates a tension between the measurements at the two colliders in specific simple models~\cite{AguilarSaavedra:2011hz}. 
This motivates the interest in a more precise measurement of the \ttbar\ production charge asymmetry.

In this paper, a measurement of the \ttbar\ production charge asymmetry in the single--lepton final state is reported. 
To allow comparisons with theory calculations, a Bayesian unfolding procedure is applied to account for distortions due to acceptance and detector effects, leading to parton--level \Ac\
measurements.
Compared with the previous \ttbar\ production charge asymmetry measurement performed by the ATLAS experiment and described in ref.~\cite{ATLAS_ljets}, the full 2011 data sample is now used 
and new differential \Ac\ measurements are performed. 
In particular, an inclusive \Ac\ measurement and measurements of \Ac\ as a function of \mttbar, \ptttbar\ and \yttbar\ are presented. 
The inclusive \Ac\ result and the differential result as a function of \mttbar\ are also presented with the additional requirement of a minimum velocity $\beta_{z,\ttbar}$ of the
\ttbar--system along the beam axis to enhance the sensitivity to BSM effects~\cite{beta_cut}. 

\section{Data sample, simulated samples and event selection}
\label{Data}
\subsection{Samples}
\label{Data and Monte Carlo simulated samples}
The measurement is performed using 7 \TeV\ $pp$ collisions recorded by the ATLAS detector~\cite{Aad:2008zzm} at the LHC during 2011.
The ATLAS detector is composed of inner tracking detectors
immersed in a 2 T axial magnetic field provided by a solenoid, surrounded by calorimeters and, as an outer layer, by a muon spectrometer 
in a magnetic field provided by three large air-core toroid magnet systems.\footnote{ATLAS uses a right--handed coordinate system with 
its origin at the nominal interaction point (IP) in the centre of the detector and the $z$--axis along the beam pipe. 
The $x$--axis points from the IP to the centre of the LHC ring, and the $y$--axis points upward. Cylindrical 
coordinates $(r,\phi)$ are used in the transverse plane, $\phi$ being the azimuthal angle around the beam pipe. 
The pseudorapidity is defined in terms of the polar angle $\theta$ as $\eta=-\ln\tan(\theta/2)$.
Transverse momentum and energy are defined as $\pt = p\sin\theta$ and
$\ET = E\sin\theta$, respectively.} After applying detector and data--quality
requirements, the recorded data corresponds to an integrated luminosity of \lumi~\cite{lumi2011}. 

Simulated \ttbar\ events are modelled using the LO multi--parton matrix--element Monte Carlo 
(MC) generator \alpgen\ ~\cite{MAN-0301-2} with the LO CTEQ6L1~\cite{cteq61} parton distribution function (PDF) for the proton.
Parton showering and the underlying event are modelled using \herwig~\cite{COR-0001-2} and \jimmy~\cite{JButterworth:1996zw} 
with the AUET2 parameter settings~\cite{Jimmy_tuning}.
The \ttbar\ sample is generated assuming a top quark mass of 172.5~\GeV\ and it is normalised to a total inclusive cross--section of $177^{+10}_{-11}$~pb 
computed at next--to--next--to--leading--order (NNLO) in QCD including resummation of next--to--next--to--leading--logarithmic (NNLL) soft gluon terms with Top++2.0~\cite{Cacciari:2012,Barn:2012,
Mitov1,Mitov2,Mitov3,Mitov4}. 
The uncertainties included in the calculation are those related to the choice of the PDF set (following the PDF4LHC prescriptions~\cite{pdf4lhc}), the variations of 
$\alpha_S$ and the choice of renormalisation and factorisation scales. These uncertainties are added in quadrature to give the quoted overall uncertainty.

Single--top events are generated using \acermc~\cite{AcerMC35} for the $t$--channel and \mcnlo\ for the $Wt$-- and $s$-- channels. The 
production of $W$ and $Z$ bosons in association with jets is simulated using 
the \alpgen \ generator interfaced to \herwig \ and \jimmy. Simulated $W$+jets events are reweighted using the NLO PDF set CT10.
Pairs of $W/Z$ bosons ($WW$, $WZ$, $ZZ$) are produced using \herwig. 

All simulated samples are generated with multiple $pp$ interactions per bunch crossing (event pile--up). Up to 24 interactions per bunch crossing were observed during the data taking period.
The number of interaction vertices in simulated samples is adjusted so that its distribution reproduces the one observed in data.
The samples are then processed through the GEANT4~\cite{AGO-0301-2} simulation~\cite{2010EPJC...70..823A} of the ATLAS detector and the same reconstruction software used for data.

\subsection{Event selection}
\label{evselection}
Candidate events with the \ttbar\ single--lepton signature are considered. These events are characterised by exactly one high--\pt\ isolated
lepton (electron, muon or tau decaying to electron or muon), missing transverse momentum \met\ due to the neutrino from the leptonic $W$ decay, two jets originating from $b$--quarks and
two jets originating from light quarks from the hadronic $W$ decay. 
 
Events are required to pass the single--electron or single--muon trigger, with thresholds in transverse energy ($E_T$) at \mbox{20 GeV} or \mbox{22 GeV} for electrons (depending on 
instantaneous luminosity conditions during the different data collection periods) and in  transverse momentum ($p_T$) at \mbox{18 GeV} for muons.
Electron candidates are required to have $E_{\rm{T}}>25$~\GeV\ and $|\eta_{\rm cluster}|<2.47$,
where $\eta_{\rm cluster}$ is the pseudorapidity of the electromagnetic energy cluster in the calorimeter. 
Candidates in the transition region $1.37<|\eta_{\rm cluster}|<1.52$ between calorimeter sections are excluded.
Muon candidates are required to have $\pt>20$~\GeV\ and $|\eta|<2.5$. 
Electrons and muons are required to be isolated to reduce the backgrounds from hadrons mimicking lepton signatures and heavy--flavour decays inside jets. 
For electrons, stringent cuts both on the shape of the calorimetric energy deposits and on the tracks used to compute the isolation, in order to reject the tracks related to photon 
conversions, are applied. Cuts that depend on $\eta$ and $E_T$ leading to a 90\% efficiency are used in a cone of $\Delta R=0.2$ for the energy isolation and in a cone of 
$\Delta R=0.3$ for the track isolation around the electron candidate.
For muons, the sum of track transverse momenta in a cone of $\Delta R=0.3$ around the muon is required to be less than 2.5~\GeV, while the total energy deposited in a cone of 
$\Delta R=0.2$ around the muon is required to be less than 4~\GeV.

Jets are reconstructed from topologically connected calorimetric energy clusters using the anti--$k_t$ algorithm~\cite{Cacciari:2008gp} with a radius parameter $R=0.4$. 
They are first calibrated to the electromagnetic energy scale and then corrected to the hadronic energy scale using 
energy-- and $\eta$--dependent correction factors obtained from simulation and control data analyses~\cite{ATL-CONF-2013-004}. The compatibility of the jets with the primary
vertex (defined as the vertex with the highest sum of the square of the transverse momenta of the tracks associated to it) is determined using the tracks associated with the jet (jet vertex fraction). 
Jets originating from the hadronisation of $b$--quarks
are identified by combining the information from three $b$--tagging algorithms, based on the topology
of $b$-- and $c$--hadron weak decays inside jets~\cite{JetFitter} and on the
transverse and longitudinal impact parameter significance of each track within the jet~\cite{SV0Tagger}.
These three tagging algorithms are combined into a single discriminating variable used to make the tagging decision.
The operating point chosen corresponds to a 70\% tagging efficiency for $b$--quarks. The rejection rate is about 150 for light--quark jets, 5 for charm jets and 14 for hadronically 
decaying $\tau$ leptons. All these numbers are evaluated in simulated \ttbar\ events.

The missing transverse momentum is reconstructed from clusters of energy deposits in the calorimeters calibrated at the electromagnetic scale and corrected according to the energy scale 
of the associated physics object. Contributions from muons are included using their momentum measured by the inner tracking and muon spectrometer systems.

Jets within $\DR = 0.2$ of an electron candidate are removed to avoid double counting electrons as jets.
Subsequently, electrons and muons within $\Delta R = 0.4$ of a jet axis and with $\pt>20$~\GeV\ are removed in order to reduce the contamination
caused by leptons from hadron decays.

In the muon channel, events are required to satisfy $\met>20\GeV$ and $\met+m_{\rm T}(W)>60\GeV$ in order to suppress the multi--jets background.\footnote{In events with a leptonic decay
of a genuine $W$ boson, $m_{\rm T}(W)$ is the $W$ boson 
transverse mass, defined as $\sqrt{ 2 \pt^{\ell} \pt^{\nu} (1-\cos ( \phi^{\ell} - \phi^{\nu} )) }$, where the measured $\met$ vector provides the neutrino information.}
In the electron channel, the multi--jets contamination is larger, and more stringent cuts 
of $\met > 30\GeV$ and $m_{\rm T}(W)>30\GeV$ are applied. 

Finally, events are required to have at least four jets with $\pt>25\GeV$ and $|\eta|<2.5$. 
These requirements define the `pretag' selection. 
For the `tag' selection, at least one of these jets is required to be $b$--tagged.

\subsection{Background estimation}
\label{Backgrounds}
The main backgrounds affecting the measurement come from $W$ bosons produced in association with jets ($W$+jets), single--top, $Z$+jets, production of $W/Z$ bosons pairs and multi--jet 
events with background leptons.\footnote{The term `background (bkgd) leptons' in this paper refers to hadrons mimicking lepton signatures and to leptons arising from heavy--hadron decays or photon
conversions.}
The $W$+jets and multi--jets contributions are evaluated using a data--driven approach. Single--top, $Z$+jets and diboson production are evaluated using simulated samples 
normalised to the approximate NNLO cross section for single--top events, NNLO cross section for inclusive $Z$ events, and NLO cross section 
for diboson events, respectively.

For reconstructed \ttbar\ candidate events, the dominant $W$+jets background is asymmetric in \DY{}and therefore a data--driven technique 
is used to estimate its normalisation.
The approach used is based on the fact that the production rate of $W^{+}$+jets is larger 
than that of $W^{-}$+jets.
Since, to a good approximation, processes other than $W$+jets give equal numbers of positively and negatively charged leptons, the formula
\begin{linenomath}
\begin{equation}
\label{chargeaform}
N_{W^+} + N_{W^-} = \left( {r_{\rm{MC}}+1 \over r_{\rm{MC}}-1} \right) (D^+ - D^-),
\end{equation}
\end{linenomath}
is used to estimate the total number of $W$ events in the selected sample, after the numbers of single--top, diboson and $Z$+jets events are evaluated in simulated samples and subtracted. 
Here, $N_{W^\pm}$ is the estimated number of $W^\pm$+jets events, $D^+ (D^-)$
is the total number of events in data passing the pretag selection described in section~\ref{evselection} 
with positively (negatively) charged leptons, and $r_{\rm{MC}}= N(pp\rightarrow W^++X )/N(pp\rightarrow W^-+X)$ is evaluated from simulation, using the \alpgen\ generator with the same 
event selection. Further details of the method can be found in ref.~\cite{ATLAS_ljets}.

The $W$ charge asymmetry depends also on the $W$+jets flavour composition, i.e. on the mixture of $Wbb$+jets, $Wcc$+jets, $Wc$+jets and $W$+light--jets processes in \alpgen\ simulated
samples. 
Since this composition cannot be predicted with sufficient precision, data--driven corrections are derived. The relative fractions are estimated in data, after subtracting all non--$W$ 
contributions, including \ttbar, applying the tag selection but requiring
the presence of exactly two jets in the final state, in order to have a control region dominated by $W$+jets events. The 
overall number of $W$+jets events is determined simultaneously with the 
heavy--flavour composition in this region. The heavy--flavour fractions in the simulated $W$+jets samples are then rescaled to the measured fractions. 
For the electron channel, the scale factors obtained are: $1.4 \pm 0.4$ for $Wbb$+jets and $Wcc$+jets, $0.7 \pm 0.4$
for $Wc$+jets and $1.00 \pm 0.10$ for $W$+light--jets components. For the muon channel, they are: $1.2 \pm 0.4$ 
for $Wbb$+jets and $Wcc$+jets, $1.0 \pm 0.4$ for $Wc$+jets and $0.97 \pm 0.09$ for $W$+light--jets components. The uncertainties include both the statistical and the systematic
components. The sources of systematic uncertainty considered are those described in section~\ref{Systematics}.

With the determined flavour fractions, the $W$+jets normalisation for pretag--selected events using eq.~\ref{chargeaform} is computed and then extrapolated to the tag--selected 
events using the tagging fractions (i.e. the fraction of events with at least one $b$--jet) computed in simulated samples. The scale factors that are applied to the tag--selected 
$W$+jets events are $0.83 \pm 0.31$ and $0.94 \pm 0.17$ in the electron and muon channel 
respectively. The uncertainties include both the statistical and the systematic components, including a particular systematic uncertainty that accounts for differences in the flavour
composition between the signal region and the region where the flavour fractions are extracted. It is derived from studies of \alpgen\ parameter variations (factorisation and
renormalisation scales, angular matching parameters and jet \pt\ generation thresholds) and it amounts to 15\% for the $Wbb/Wcc/Wc$+jets components and 5\% for the $W$+light--jets component.

The `Matrix Method' is used to evaluate the multi--jets background with background leptons. The method relies on defining `loose' and 
`tight' lepton samples~\cite{Aad:2010ey} and measuring the `tight' selection efficiencies for real ($\epsilon_\mathrm{real}$) and background ($\epsilon_\mathrm{bkgd}$) `loose' leptons. The `loose' selection requires less stringent
identification and isolation requirements than the ones described in section~\ref{evselection}, referred here as `tight' selection.
The fraction $\epsilon_\mathrm{real}$ is measured using data control samples of $Z$ boson decays to two 
leptons. The fraction  $\epsilon_\mathrm{bkgd}$ is measured in control regions where the contribution of background leptons is dominant. 

The expected and observed yields are listed in table~\ref{evtnumbers}. The number of events in the electron channel
is significantly lower than in the muon channel due to the higher lepton \pt\ threshold, tighter isolation and the more stringent missing transverse momentum requirements. The number of
events observed in data and the total predicted yield are compatible within uncertainty.
\begin{table*}[tbp]
\begin{center}
{\footnotesize
  \begin{tabular}{ |l | rrr | rrr | rrr | rrr |}
      \hline
      Channel & \multicolumn{3}{|c|}{$\mu$ + jets pretag} & \multicolumn{3}{|c|}{$\mu$ + jets tag} & \multicolumn{3}{|c|}{$e$ + jets pretag} & \multicolumn{3}{|c|}{$e$ + jets tag} \\ 
      \hline
%     $t\bar{t}$       & 32800 &$\!\!\!\pm\!\!\!$& 3500  & 28400 &$\!\!\!\pm\!\!\!$& 3100  & 20100 &$\!\!\!\pm\!\!\!$& 2200  & 17400 &$\!\!\!\pm\!\!\!$& 1900  \\
      $t\bar{t}$       & 34900 &$\!\!\!\pm\!\!\!$& 2200  & 30100 &$\!\!\!\pm\!\!\!$& 1900  & 21400 &$\!\!\!\pm\!\!\!$& 1300  & 18500 &$\!\!\!\pm\!\!\!$& 1100  \\ %NEW tt x-sec
      $W$+jets         & 28200 &$\!\!\!\pm\!\!\!$& 3100  & 4800  &$\!\!\!\pm\!\!\!$& 900   & 13200 &$\!\!\!\pm\!\!\!$& 1600  & 2300  &$\!\!\!\pm\!\!\!$& 900  \\
      Multi--jets      & 5500  &$\!\!\!\pm\!\!\!$& 1100  & 1800  &$\!\!\!\pm\!\!\!$& 400   & 3800  &$\!\!\!\pm\!\!\!$& 1900  & 800   &$\!\!\!\pm\!\!\!$& 400  \\ 
      Single top       & 2460  &$\!\!\!\pm\!\!\!$& 120   & 1970  &$\!\!\!\pm\!\!\!$& 100   & 1530  &$\!\!\!\pm\!\!\!$& 80    & 1220  &$\!\!\!\pm\!\!\!$& 60   \\ 
      $Z$+jets         & 3000  &$\!\!\!\pm\!\!\!$& 1900  & 480   &$\!\!\!\pm\!\!\!$& 230   & 3000  &$\!\!\!\pm\!\!\!$& 1400  & 460   &$\!\!\!\pm\!\!\!$& 220   \\
      Diboson          & 380   &$\!\!\!\pm\!\!\!$& 180   & 80    &$\!\!\!\pm\!\!\!$& 40    & 230   &$\!\!\!\pm\!\!\!$& 110   & 47    &$\!\!\!\pm\!\!\!$& 22   \\ 
      Total background & 40000 &$\!\!\!\pm\!\!\!$& 4000  & 9200  &$\!\!\!\pm\!\!\!$& 1000  & 21700 &$\!\!\!\pm\!\!\!$& 2900  & 4800  &$\!\!\!\pm\!\!\!$& 1000  \\ 
      Signal + background     & 74000 &$\!\!\!\pm\!\!\!$ & 4000 & 39300 &$\!\!\!\pm\!\!\!$& 2100  & 43100 &$\!\!\!\pm\!\!\!$& 3100 & 23300 &$\!\!\!\pm\!\!\!$& 1600  \\ \hline
      Observed                & \multicolumn{3}{|c|}{70845}  & \multicolumn{3}{|c|}{37568} & \multicolumn{3}{|c|}{40972}      & \multicolumn{3}{|c|}{21929} \\
      \hline
    \end{tabular}
    }
\caption{\label{evtnumbers}Numbers of expected events for the \ttbar\ signal and the various background processes and observed events in data for the pretag and tag samples. The uncertainties include statistical and systematic components.}
\end{center}
\end{table*}

\section{The \ttbar\ production charge asymmetry measurement}
\label{Measurement}
After the reconstruction of the \ttbar--system 
(section~\ref{Reconstruction}) and the estimation of the background, the \DY{}spectra (section~\ref{Unfolding}) are unfolded to obtain inclusive and differential parton--level charge 
asymmetry measurements (as a function of \mttbar, \ptttbar\ and \yttbar), as defined in eq.~\ref{Ac}.

In addition, an inclusive measurement and a differential measurement as a function of \mttbar\ are performed for events where the $z$--component of the \ttbar--system 
velocity is large, $\beta_{z,\ttbar}>0.6$. Most BSM models introduced to explain the excesses in the CDF and D0 measurements postulate the presence of new particles that can alter the SM 
prediction for \Ac. Requiring $\beta_{z,\ttbar}>0.6$ defines a region of phase--space where the effects of these new particles on the asymmetry are enhanced~\cite{beta_cut}. 
\subsection{Reconstruction of the \ttbar--system}
\label{Reconstruction}
A kinematic fit is used to determine the likelihood for candidate events to be \ttbar\ events as well as to 
determine the four--vector of the top quark and antiquark to compute \DY{}. The charge of the lepton is used to determine whether the reconstructed object is a top
quark or antiquark. A detailed description of the method and its assumptions can be found in ref.~\cite{ATLAS_ljets}. 
In simulation studies using \ttbar\ events, the fraction of events reconstructed with the correct \DY{}sign was evaluated to be 75\%.

\begin{figure*}[!htb]
  \begin{center}
 \includegraphics[width=0.41\textwidth]{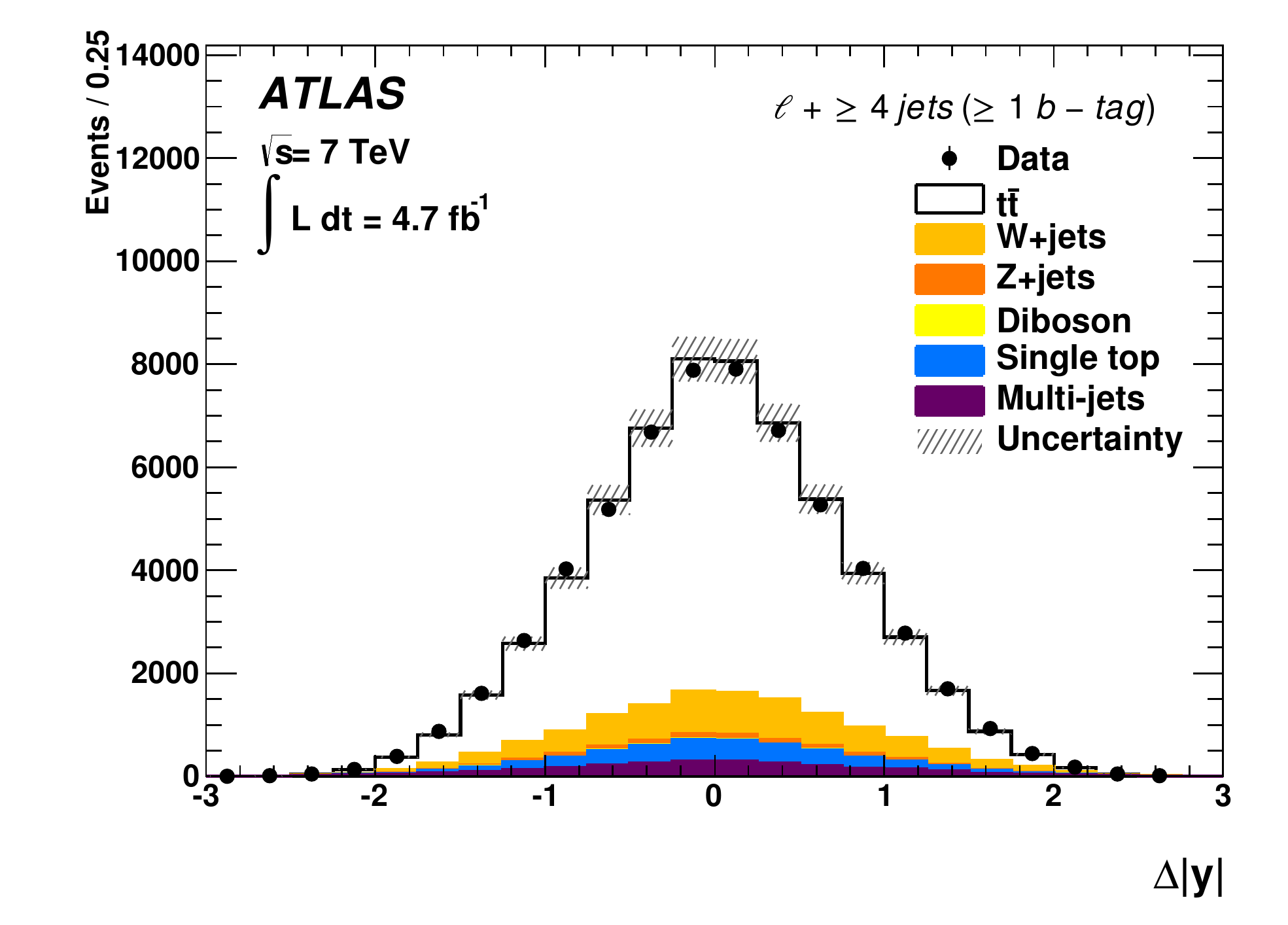}
 \includegraphics[width=0.41\textwidth]{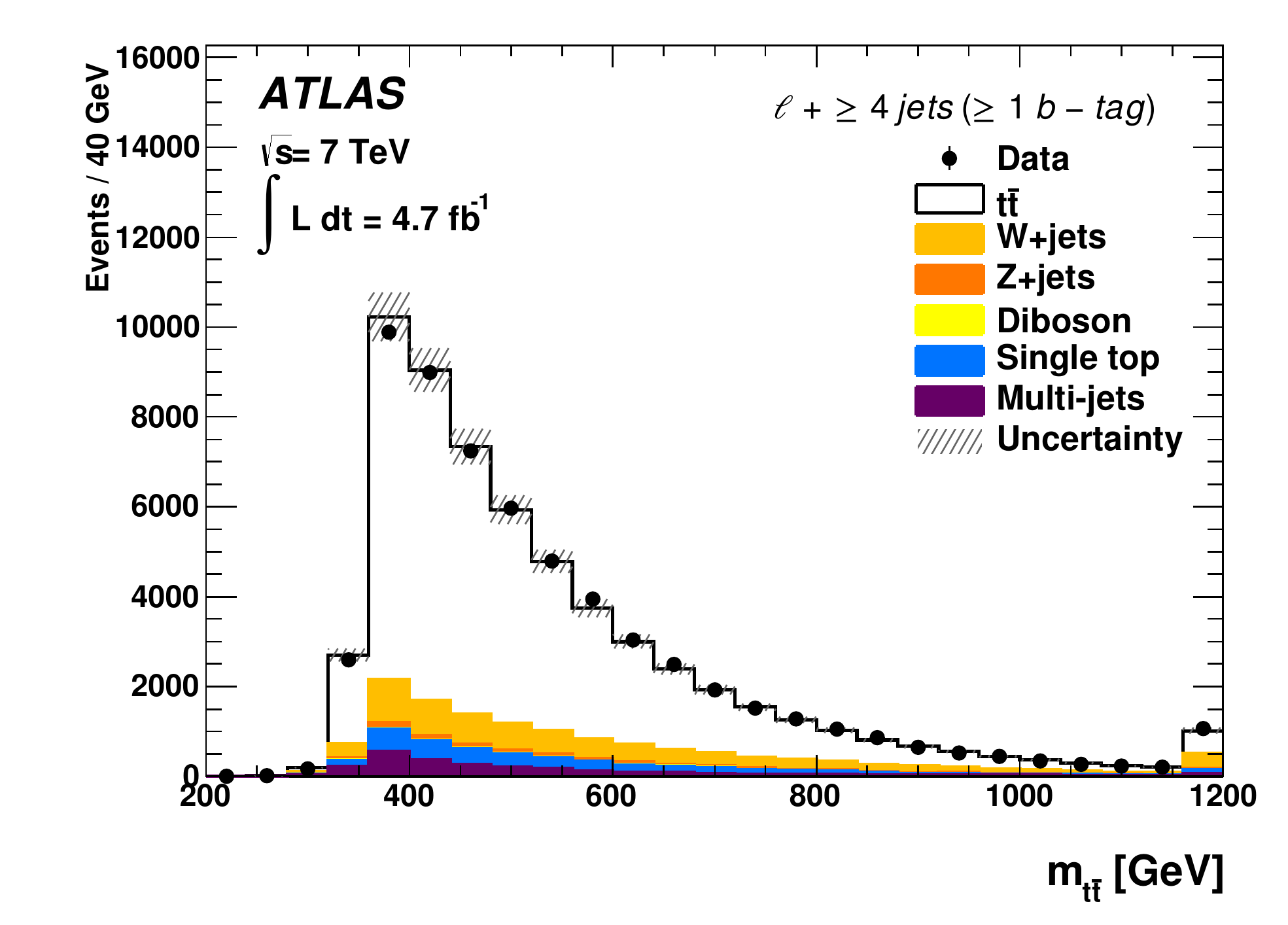}
 \includegraphics[width=0.41\textwidth]{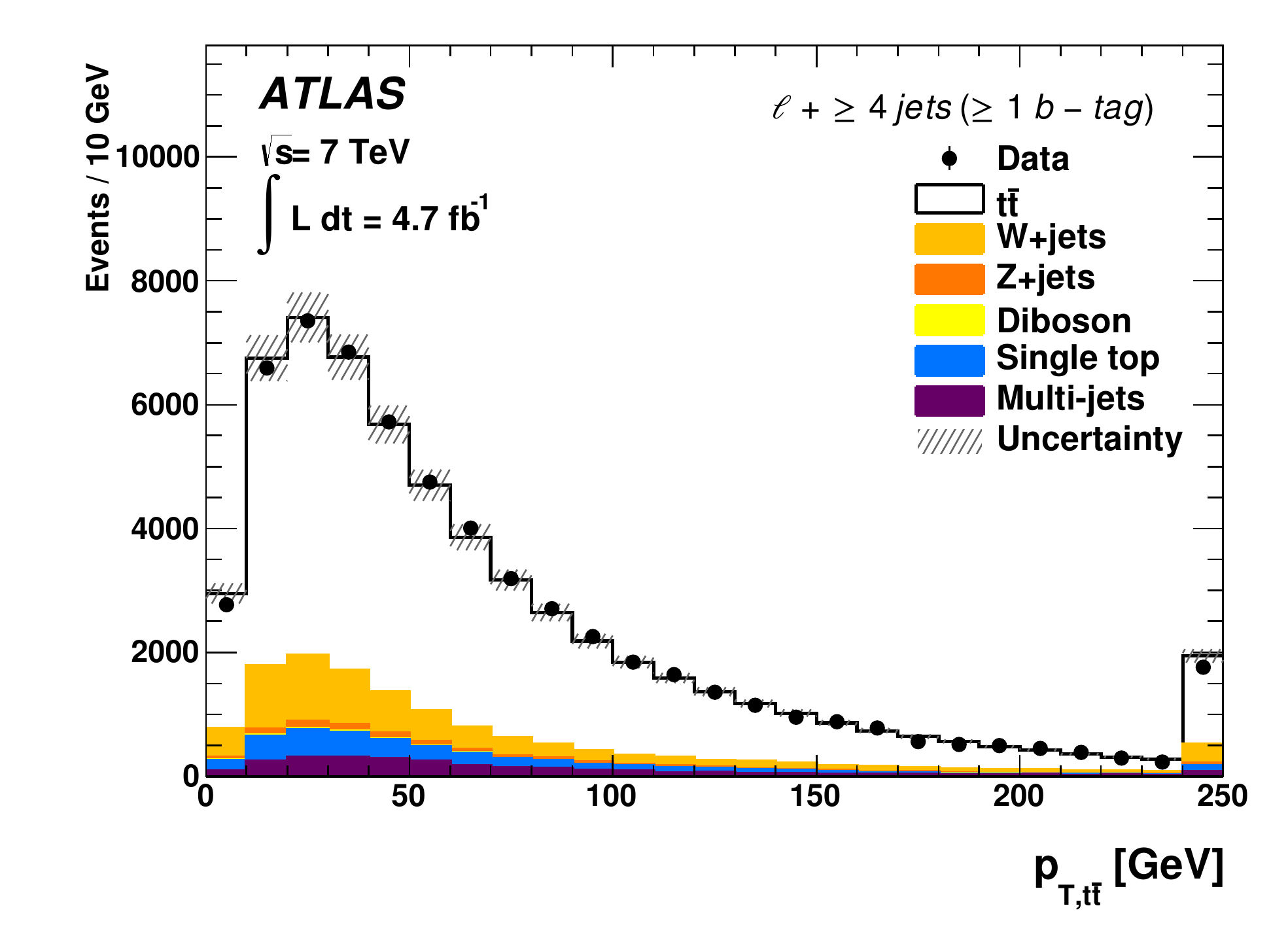}
 \includegraphics[width=0.41\textwidth]{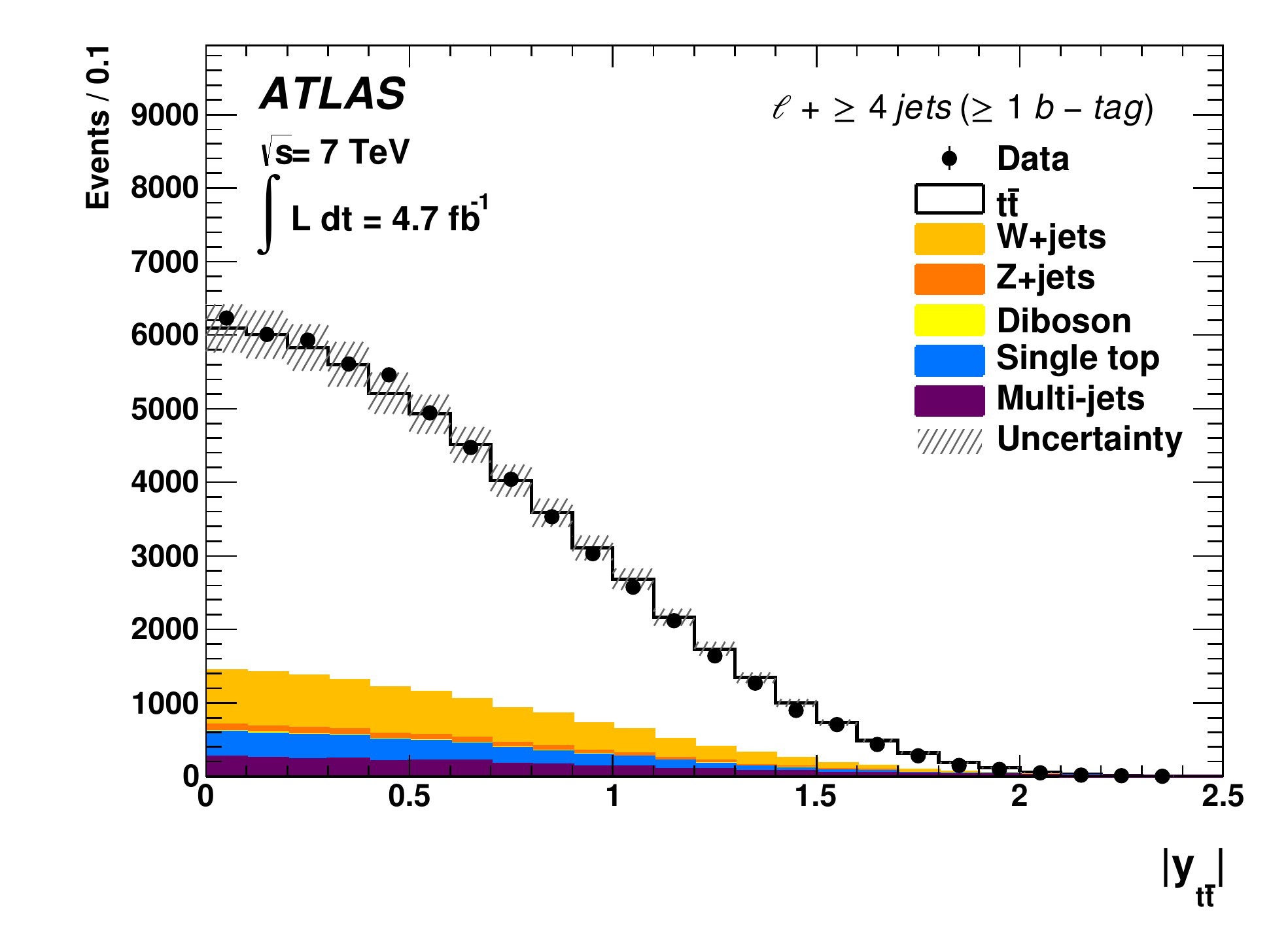}
 \includegraphics[width=0.41\textwidth]{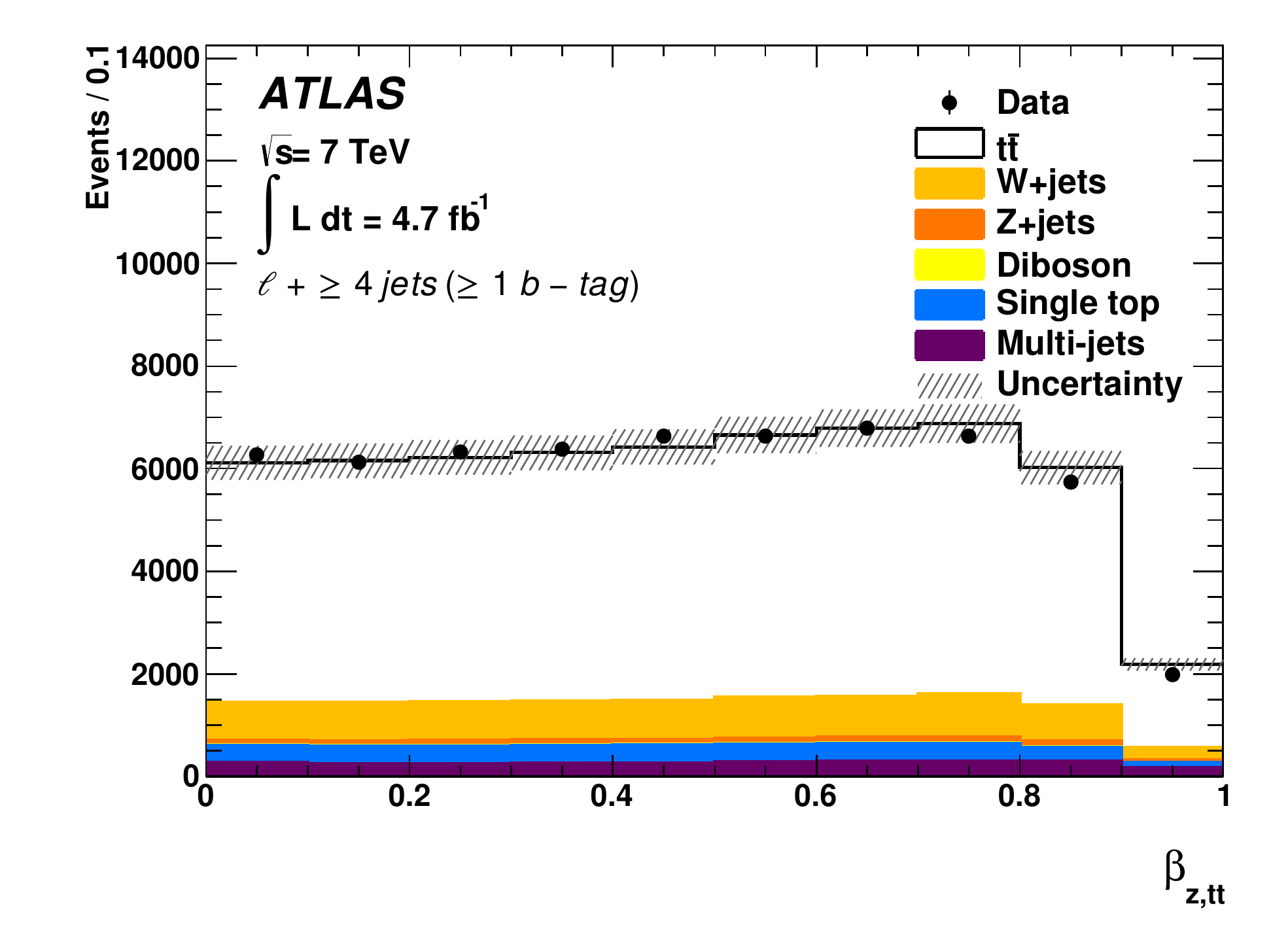}
    \caption{Reconstructed \DY\ (top left), invariant mass \mttbar\ (top right), transverse momentum \ptttbar\ (centre left), rapidity \yttbar\ (centre right) and velocity
    $\beta_{z,\ttbar}$ (bottom) distributions for the electron and muon channels combined after requiring at least one $b$--tagged jet. Data (dots) and SM expectations (solid lines) are shown. 
The uncertainty on the total prediction includes both the statistical and the systematic components.
The overflow is included in the last bin.}
    \label{fig:DYtop}
  \end{center}
\end{figure*}

For the differential measurements a cut on the likelihood is applied to reject badly reconstructed events, reducing the migrations across the bins. 
The reconstructed \DY{}distribution is shown in figure~\ref{fig:DYtop} along with the distributions of \mttbar, \ptttbar, \yttbar\ and $\beta_{z,\ttbar}$. 

\subsection{Unfolding procedure}
\label{Unfolding}
The reconstructed \DY{}distributions are
distorted by acceptance and detector resolution effects.
We use the Fully Bayesian Unfolding (FBU)~\cite{Fbu2012arXiv1201.4612C} technique to estimate
the parton--level distributions from the measured spectra.
This method relies on applying Bayes' theorem to the unfolding problem, which can be formulated in the following terms.

Given an observed data spectrum $\Data\in\Integer^{N_{\rm{r}}}$ and a migration matrix
$\TrasfMatrix\in\Real{}^{N_{\rm{r}}}\times{}\Real{}^{N_{\rm{t}}}$ ($N_{\rm{r}}$ and $N_{\rm{t}}$ are the number of bins in the measured and true spectra respectively) that takes into account the distortion effects 
mentioned above, the posterior probability density of the true spectrum $\Truth{}\in\Real{}^{N_{\rm{t}}}$ follows the probability density
\begin{linenomath}
\begin{equation}
\conditionalProb{\Truth{}}{\Data{},\TrasfMatrix{}}
\propto{}
\conditionalLhood{\Data{}}{\Truth{},\TrasfMatrix{}}
\cdot{}
\pi{}\left(\Truth{}\right)
\nonumber
\end{equation}
\end{linenomath}
where \conditionalLhood{\Data{}}{\Truth{},\TrasfMatrix{}} is the conditional likelihood for the data $\Data{}$ assuming the true $\Truth{}$ and the migration matrix $\TrasfMatrix{}$, and 
$\pi{}$ is the prior probability density for the true $\Truth{}$.

Assuming that the data follows a Poisson distribution, the likelihood
\conditionalLhood{\Data{}}{\Truth{},\TrasfMatrix{}} can be computed starting from the
migration matrix \TrasfMatrix{}, whose elements $\TrasfMatrix{}_{tr}$ represent the probability and the efficiency of an event produced in the true bin $t$ to be reconstructed in any bin
$r$. The background in each bin is taken into account when computing \conditionalLhood{\Data{}}{\Truth{},\TrasfMatrix{}}. 
While the above quantities can be estimated from
simulated samples of signal events, the prior probability density $\pi{}(\Truth{})$ must be chosen according to 
what is known about $\Truth{}$ before the measurement. In this context, the
choice of the prior can be interpreted as the choice of a regularisation in other unfolding techniques
(see ref.~\cite{D'Agostini:1994zf} for instance).
After choosing a prior, the posterior probability density $\conditionalProb{\Truth{}}{\Data{},\TrasfMatrix{}}$ is computed by 
generating uniformly distributed points in the
$N_{\rm{t}}$--dimensional space, and evaluating for each of them 
\conditionalLhood{\Data{}}{\Truth{},\TrasfMatrix{}} and $\pi{}(\Truth{})$. 
A weight given by \mbox{$\conditionalLhood{\Data{}}{\Truth{},\TrasfMatrix{}}\cdot{}\pi(\Truth{})$} is then assigned to each
point, allowing the posterior
probability density of the unfolded spectrum to be determined, for each \DY{}bin and for \Ac.

The FBU method has two main advantages. Firstly, it gives a precise physical meaning to the regularisation procedure through the choice of a prior built with well-motivated physical 
quantities. Secondly, systematic uncertainties are accounted for consistently with the Bayesian statistical approach, by reporting credible intervals built by integrating the posterior 
distribution over the nuisance parameters.

The choice of the prior is arbitrary. With a flat prior, the FBU method has been checked to be equivalent to unregularised matrix inversion.
Non--uniform priors favour spectra that have some well--defined features.
By assuming that some spectra are more likely than others, information is added to the measurement, reducing the uncertainty but potentially biasing its outcome. 

Two different priors are used in the following: a flat prior and a curvature
prior. The curvature prior is defined starting from the definition of the curvature
$C(\Truth{})$ being the sum of the squares of the second derivatives of the \DY{}distribution \Truth{} with
$N_{\rm{t}}$ bins:
\begin{linenomath}
\begin{equation}
\label{curvature}
C(\Truth{}) = \sum\limits_{i=2}^{N_{\rm{t}}-1} (\Delta_{i+1,i} - \Delta_{i,i-1})^2 ,\\
\nonumber
\end{equation}
\end{linenomath}
where $\Delta_{a,b}=T_{a}-T_{b}$. The curvature prior is then defined as follows:
\begin{linenomath}
\begin{equation}
\prior{}
\propto{}
\begin{cases}
%e^{-\alpha{}S(\Truth{})} & \text{if } T_{t}\in \text{integration space}, \forall{}t\in[1, N_{\rm{t}}] \\
\rm{e}^{\alpha{}S(\Truth{})} & \text{in the integration space}, \forall{}t\in[1, N_{\rm{t}}] \\
0 & \text{otherwise}
\end{cases}
\label{eq:curv_prior}
\end{equation}
\end{linenomath}
where $\alpha$ is the regularisation parameter and \ST{} is a regularisation function, 
defined, for each generated point, as the difference between the curvature $C(\Truth{})$ of the true \DY{}spectrum \Truth{} and that of the estimated spectrum $\Truth{}^*$.

The flat prior is used for the differential measurements of \Ac\ as a
function of \mttbar\ and of \yttbar{}.  The curvature prior defined in
eq.~\ref{eq:curv_prior} is used for the inclusive measurement and for
the differential measurement as a function of \ptttbar{}, because it
reduces the uncertainty on these measurements.  The regularisation
strength $\alpha = 10^{-8}$ is chosen based on the numerical value of
the curvature of the true spectrum. It has been checked, by varying $\alpha$
by one order of magnitude included the $\alpha=0$ unregularised
case, that this particular choice of $\alpha$ does not cause any
significant bias in either the unfolded distributions or in the
computed asymmetries. The consistency of the
FBU method with the iterative scheme~\cite{D'Agostini:1994zf} has been checked as well.

Four bins are used for the \DY{} distribution both for the inclusive and the differential measurements. The \DY{} bin ranges are the same in both measurements. The bin ranges for the 
differential variables are chosen to have approximately the same number of entries in each bin. The \Ac\ posterior probability density is built from the asymmetry in each
generated point of the integration space. The value of \Ac\ and its statistical uncertainty are the mean and the RMS of the posterior probability density distribution respectively.

\subsection{Systematic uncertainties}
\label{Systematics}
Several sources of systematic uncertainty are taken into account. 

A possible small mis--modelling of the lepton momentum scale and resolution in simulation is corrected by scale factors derived
from the comparison of $Z\rightarrow\ell\ell$, $J/\psi\rightarrow\ell\ell$ and $W\rightarrow e\nu$ events in data and simulation.
The uncertainty on the scale factors ranges from 1\% to 1.5\% depending on the \pt\ and $\eta$ of the leptons.

The jet energy scale is derived using information from test--beam
data, collision data and simulation. Its uncertainty is between 1\% and 2.5\% in the central region of the detector, depending on jet \pt\ and
$\eta$~\cite{ATL-CONF-2013-004}.
This value includes uncertainties due to the flavour composition of the sample, mis--measurements due to the effect of nearby jets, influence of pile--up, and a \pt--dependent uncertainty
for jets arising from the fragmentation of $b$--quarks.
The jet energy resolution and reconstruction efficiencies are measured in data using
techniques described in refs.~\cite{ATL-CONF-2013-004,ATLAS-CONF-2010-054}. 

The uncertainties on the lepton and jets are propagated to the missing transverse momentum calculation.

The $b$--tagging efficiencies and light jets mis--tag rates are measured in data. Jet \pt--dependent scale factors are applied to
simulation to match the efficiencies observed in data. The typical uncertainty on the $b$--tagging scale factors ranges from 6\% to 20\% (depending on jet \pt\ and $\eta$) for $b$--jets,
from 12\% to 22\% for $c$--jets and is about 16\% for light--jets~\cite{SV0Tagger}. The impact of this uncertainty is negligible.

The systematic uncertainty in the modelling of the signal process 
is assessed by varying the simulation parameters and by using a different Monte Carlo generator (\powheg~\cite{nason:2004rx,Frixione:2007vw}). 
The sources of systematic uncertainty considered
are the choice and the functional form of factorisation scale and the choice of parton shower model (\pythia\ or \herwig). 
The impact of the choice of PDFs is evaluated following the procedure described in
ref.~\cite{pdf4lhc}. All these uncertainties have a negligible impact on the asymmetry.

The limited size of the MC simulation samples gives rise to a systematic uncertainty
in the response matrix. This is estimated by independently 
varying the bin content of the response matrix according to Poisson distributions.

Several other sources of systematic uncertainties are considered, namely the uncertainties on: the luminosity determination (1.8\%)~\cite{lumi2011}, 
the lepton and trigger reconstruction and identification scale factors, the lepton charge mis--identification, the jet vertex fraction scale factor, the missing transverse momentum scale and resolution and the $Z$+jets and 
multi--jets background normalisations. All of these lead to uncertainties on the asymmetry measurements below 0.001 and are therefore negligible.

Systematic uncertainties related to the different choice of PDFs and to the shape of the $W$+jets distributions are also considered. The former is evaluated 
as explained above. The latter is estimated in simulated events generated with the same variations of the \alpgen\ parameters as described above for the modelling of the signal process. 

For each of the systematic uncertainties (except for those related to the modelling of the \ttbar\ signal and for the $W$+jets shape) the $W$+jets normalisation and the
heavy--flavour composition are 
recomputed as described in section~\ref{Backgrounds} to take into account the correlation with the various sources of systematic uncertainty considered.

For the systematic uncertainties affecting the background, the posterior probability density with a modified background prediction is computed. For
those affecting the signal, the posterior probability density with the modified efficiency and response matrix is evaluated.
 
Systematic uncertainties are taken into account with a
marginalisation procedure. After computing the posterior probability density corresponding to each systematic variation, the likelihood used in the unfolding is marginalised by integrating out its
dependence on the nuisance parameters. It is assumed that the priors for
all nuisance parameters are Gaussian and that there is no correlation
between them. A marginalisation is then performed by transforming the
integral over the nuisance parameter into a discrete sum of the
posterior probability densities evaluated at three values of the
nuisance parameter: the central one and the 1$\sigma$ variations.
The resulting posterior probability density is finally used to extract the systematic uncertainty on the measurements.

\section{Results}
\subsection{Inclusive and differential measurements}
\label{Results}
The \ttbar\ production charge asymmetry is measured to be \resultincl\, compatible with the SM prediction $\Ac=0.0123\pm0.0005$~\cite{Bern:2012}.
These values are shown in table~\ref{tab:results} together with the measurement and prediction for $\mttbar >600$ \GeV. 
The total systematic uncertainty is computed with the marginalisation procedure described in section~\ref{Systematics}. The uncertainties quoted for all the results in this section 
include statistical and systematic components.
\begin{table*}[t]
\centering
{\small
\begin{tabular}{|l| c |c|}
\hline
 \Ac & Data & Theory\\
\hline
 Unfolded & 0.006$\pm$0.010 & 0.0123$\pm$0.0005 \\
 Unfolded with $\mttbar > 600$ GeV & 0.018$\pm$0.022 & $0.0175^{+0.0005}_{-0.0004}$\\
 Unfolded with $\beta_{z,\ttbar}>0.6$ & 0.011$\pm$0.018 & $0.020^{+0.006}_{-0.007}$\\
\hline
\end{tabular}
}
\caption{\label{tab:results}Measured inclusive charge asymmetry, \Ac, values for the electron and muon channels combined after
unfolding without and with the $\beta_{z,\ttbar} > 0.6$ cut explained in the text. The \Ac\ measurement with a cut on the \ttbar\ invariant mass $\mttbar > 600 $ GeV is also shown.
SM predictions, as described in the text, are also reported.
The quoted uncertainties include statistical and systematic components after the marginalisation.}
\end{table*}
In order to estimate the impact of each source of systematic uncertainty, the marginalisation procedure is repeated removing 
one such source at a time from the global marginalisation. For each of the systematic uncertainties considered in this analysis and for all the measurements, the impact on the \Ac\ 
value and its uncertainty is less than 10\% of the statistical uncertainty, and thus negligible.

\begin{figure*}[tbp]
\begin{center}
 \includegraphics[width=0.41\textwidth]{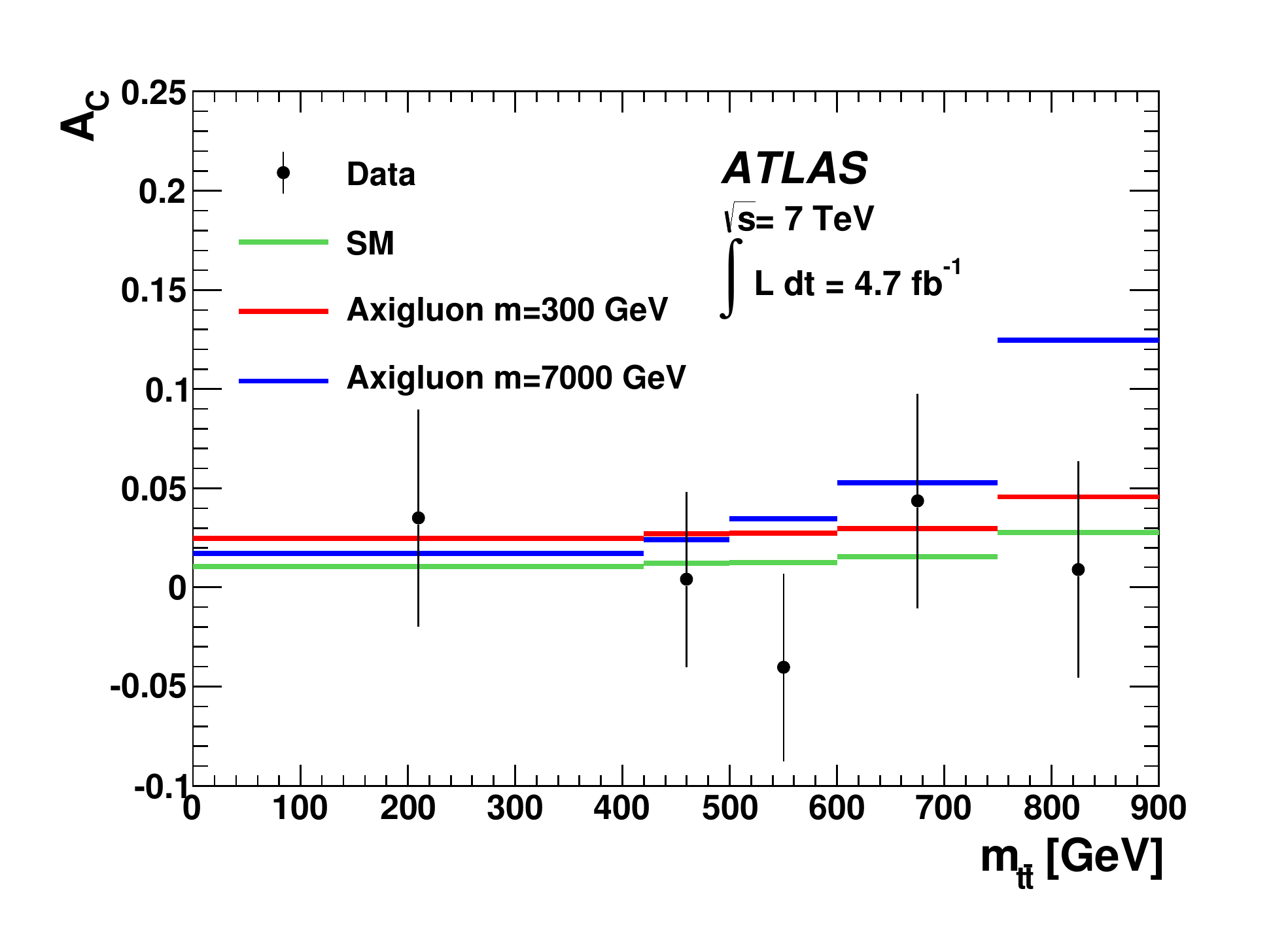}
 \includegraphics[width=0.41\textwidth]{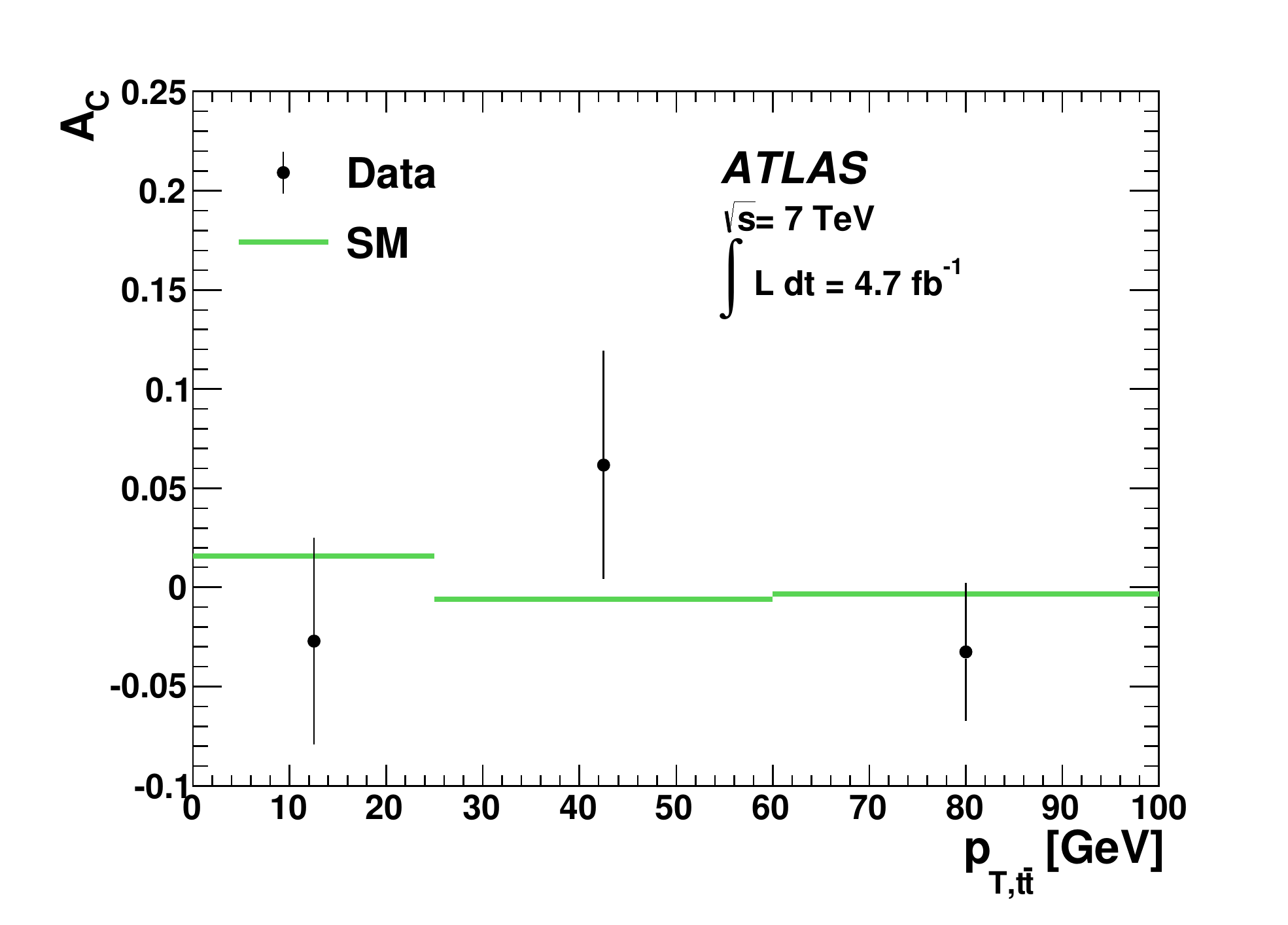}
 \includegraphics[width=0.41\textwidth]{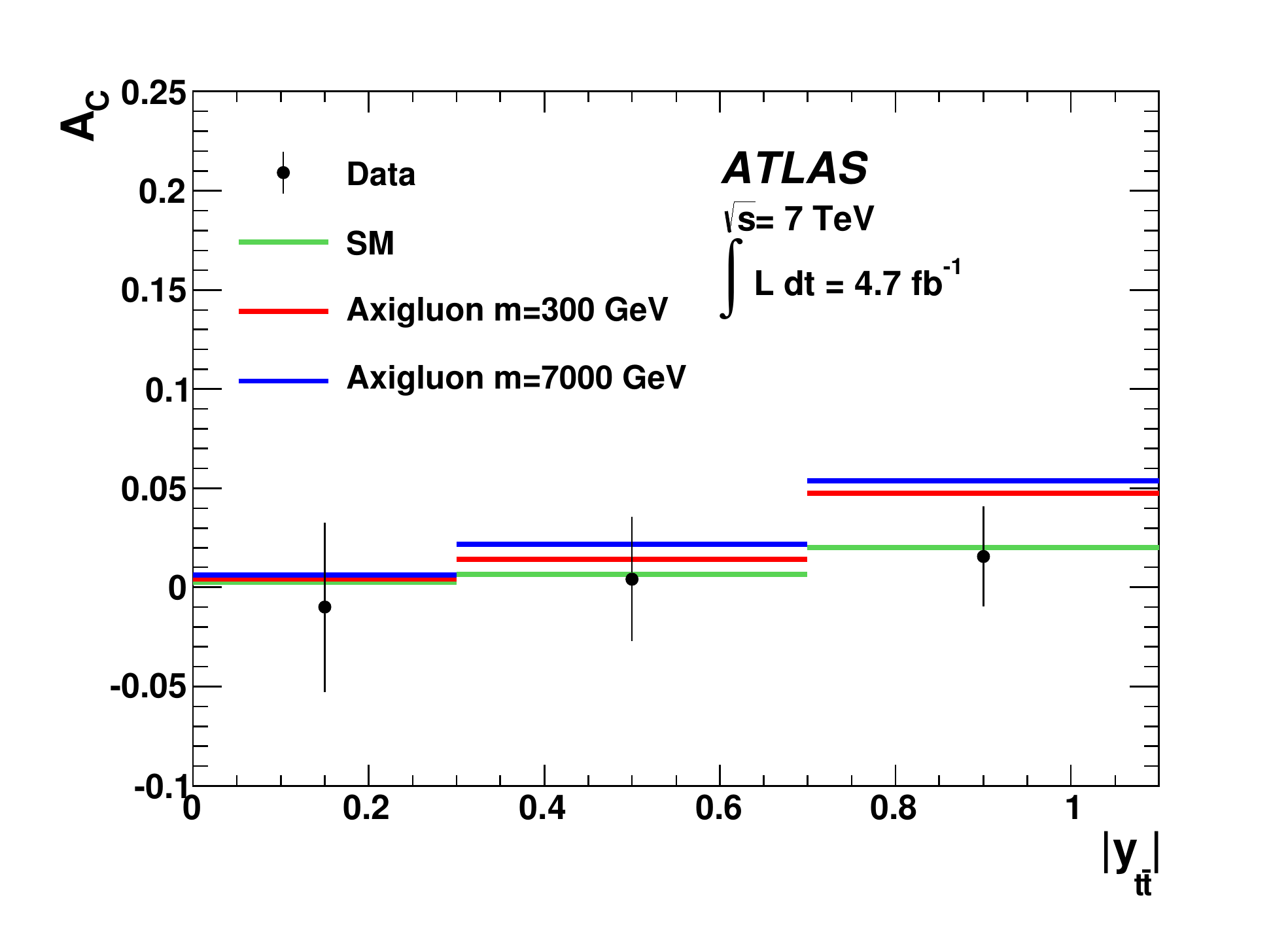}
 \includegraphics[width=0.41\textwidth]{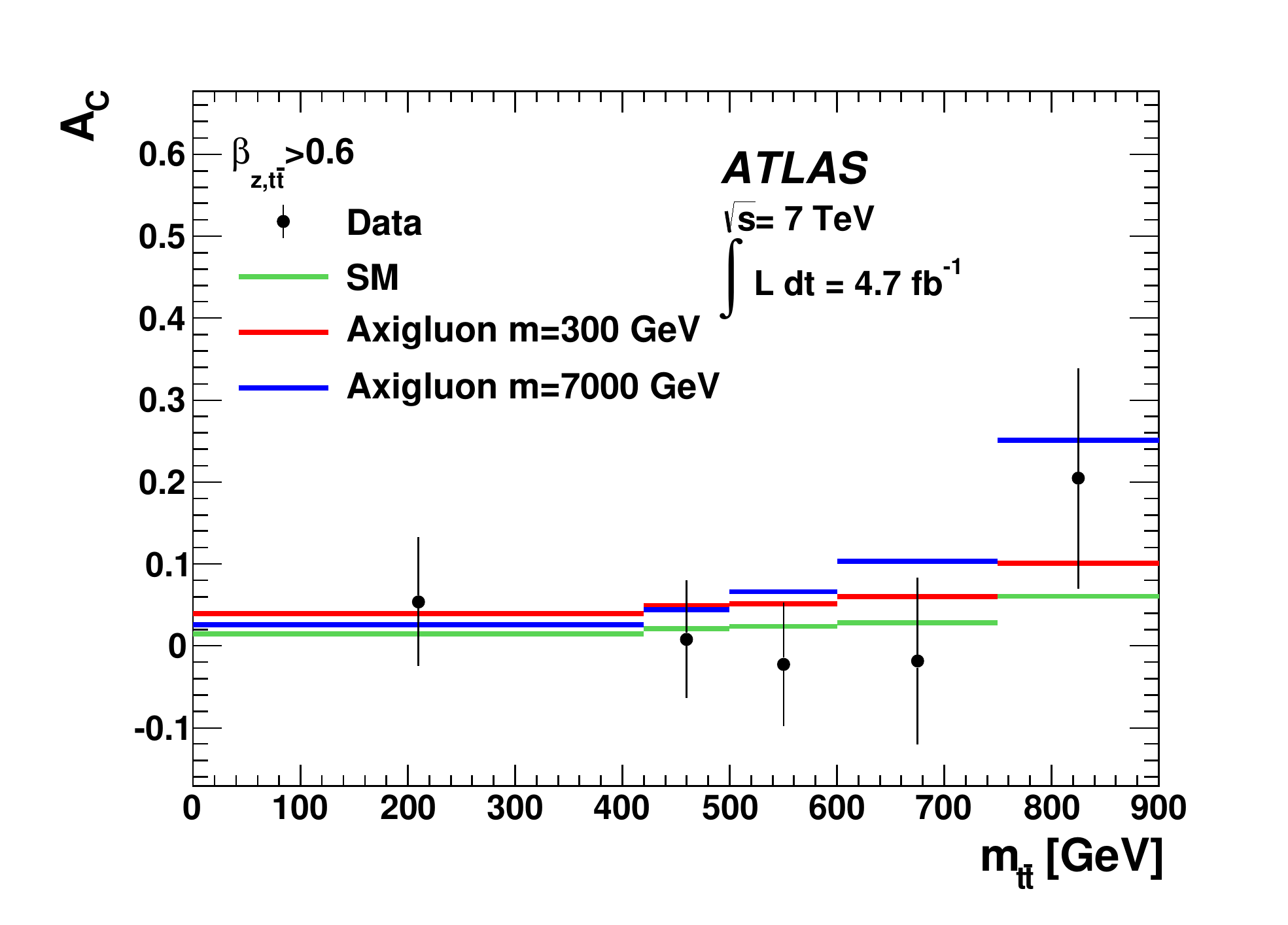}
    \caption{Distributions of \Ac\ as a function of \mttbar\ (top left), \ptttbar\ (top right) and \yttbar\ (bottom left) 
    after unfolding, for the electron and muon channels combined. The \Ac\ distribution as a function of \mttbar, after the $\beta_{z,\ttbar} > 0.6$ requirement, is also shown (bottom right).
    The \Ac\ values after the unfolding (points) are compared with the SM predictions (green lines) and the predictions for a colour--octet axigluon with a mass of 300 \GeV\ (red
    lines) and 7000 \GeV\ (blue lines) respectively, as described in the text. The thickness of the lines represents the factorisation and renormalisation
    scale uncertainties on the corresponding theoretical predictions. The values plotted are the average \Ac\ in each bin. The error bars include both the statistical and the systematic 
    uncertainties on \Ac\ values. The bins are the same as the ones reported in  tables~\ref{tab:results_diff} and ~\ref{tab:results_mtt_beta} respectively.}
    \label{fig:DY_unf_diff}
  \end{center}
\end{figure*}

As a cross--check, the systematic uncertainties affecting \Ac\ are computed one by one before the marginalisation procedure described above. For each source, the systematic 
uncertainty represents the variation of the mean of posterior probability densities corresponding to a 1$\sigma$ variation of the nuisance parameter. 
The statistical uncertainty still dominates the variations in \Ac\ even before the marginalisation procedure. Table~\ref{Table_Systematics} summarises the result of
this `cross--check' procedure for the inclusive charge asymmetry measurement (left column) and for the measurement with the $\mttbar >600$ \GeV\ requirement after unfolding (central
column).
\begin{table*}[tbp]
%\begin{center}
\centering
{\footnotesize
\begin{tabular}{|l| c| c| c|}
\hline
Source of systematic uncertainty  & \multicolumn{3}{|c|}{$\delta{\Ac}$} \\
\hline
  & Inclusive & $\mttbar >600$ \GeV & $\beta_{z,\ttbar} > 0.6$ \\
\hline
Lepton reconstruction/identification		   & $<0.001$ & 0.001    & $<0.001$\\
Lepton energy scale and resolution		   & 0.003    & 0.003    & 0.003\\
Jet energy scale and resolution		   	   & 0.003    & 0.003    & 0.005\\
Missing transverse momentum and pile--up modelling & 0.002    & 0.002    & 0.004\\
Multi--jets background normalisation		   & $<0.001$ & 0.001    & 0.001\\
$b$--tagging/mis--tag efficiency		   & $<0.001$ & 0.001    & 0.001\\
Signal modelling				   & $<0.001$ & $<0.001$ & $<0.001$\\
Parton shower/hadronisation			   & $<0.001$ & $<0.001$ & $<0.001$\\
Monte Carlo statistics				   & 0.002    & $<0.001$ & $<0.001$\\
PDF						   & 0.001    & $<0.001$ & $<0.001$\\
$W$+jets normalisation and shape          	   & 0.002    & $<0.001$ & $<0.001$\\
\hline
Statistical uncertainty				   & 0.010 & 0.021 & 0.017 \\
\hline 
\end{tabular}
}
\caption{\label{Table_Systematics}Systematic uncertainties for the inclusive asymmetry, \Ac\ (second column), the asymmetry for  $\mttbar > 600$ \GeV (third column) and the inclusive
asymmetry, \Ac, for $\beta_{z,\ttbar} > 0.6$ (fourth column).
For variations resulting in asymmetric uncertainties, the average absolute deviation from the nominal value is reported. The values reported for each systematic uncertainty are the
variation of the mean of posteriors computed considering 1$\sigma$ variations.}
%\end{center}
\end{table*}
Figure~\ref{fig:DY_unf_diff} shows the charge asymmetry as a function of \mttbar, \ptttbar\ and \yttbar\ compared with the theoretical SM predictions described in ref.~\cite{Bern:2012} and 
provided by its authors for the chosen bins. In addition, predictions for two assumed mass values (300 \GeV~\cite{AguilarSaavedra:2011ci} and 7000 \GeV), for a heavy axigluon exchanged in the 
$s$--channel, are also shown. The masses are chosen as benchmarks, taking into account the fact that they would not be visible as 
resonances in the \mttbar\ spectrum. The parameters of the model are tuned to give a forward--backward asymmetry compatible with the Tevatron results. 
The differential distributions and respective asymmetries do not show any significant deviation from the SM prediction. 
The resulting charge asymmetry \Ac\ is shown in table~\ref{tab:results_diff} for the differential measurements 
as a function of \mttbar, \ptttbar\ and \yttbar. The systematic uncertainties, computed before the marginalisation procedure as described above in the cross--check procedure, are listed 
in table~\ref{table:Systematics_diff} for each of the differential measurements.  
The correlation matrices for the statistical uncertainties are shown in table~\ref{table:varAC_diff} for the measurement as a function of \mttbar, \ptttbar\ and \yttbar\ respectively.
\begin{table*}[tbp]
\begin{center}
{\small
\begin{tabular}{|l| c |c |c| c| c| }
\hline
 &\multicolumn{5}{|c|}{$\mttbar{}$ ~[\GeV]}    \\
\hline
\Ac & $0$--$420$ & $420$--$500$ & $500$--$600$ & $600$--$750$ & $>750$ \\
\hline
Unfolded & $0.036\pm0.055$ & $0.003\pm0.044$ & $-0.039\pm0.047$ & $0.044\pm0.054$ & $0.011\pm0.054$ \\
Theory & $0.0103^{+0.0003}_{-0.0004}$ & $0.0123^{+0.0006}_{-0.0003}$ & $0.0125\pm0.0002$ & $0.0156^{+0.0007}_{-0.0009}$ & $0.0276^{+0.0004}_{-0.0008}$ \\
\hline
\end{tabular}
}
\end{center}
%\newline
\vspace*{2pt}
%\newline
\begin{center}
{\small
\begin{tabular}{|l |c |c |c |}
\hline
 &\multicolumn{3}{|c|}{$\ptttbar{}$ ~[\GeV]}    \\
\hline
\Ac &     $0$--$25$           &        $25$--$60$         &  $>60$            \\
\hline
Unfolded  &    $-0.032\pm0.052$       &     $0.067\pm0.057$       &      $-0.034\pm0.034$     \\
Theory & $0.0160^{+0.0007}_{-0.0009}$ & $-0.0058^{+0.0004}_{-0.0004}$ & $-0.0032^{+0.0002}_{-0.0002}$  \\
\hline
\end{tabular}
}
\end{center}
%\newline
\vspace*{2pt}
%\newline
\begin{center}
{\small
\begin{tabular}{|l |c |c |c |}
\hline
 &\multicolumn{3}{|c|}{$\yttbar{}$}    \\
\hline
\Ac  &     $0$--$0.3$          &        $0.3$--$0.7$     &  $>0.7$            \\
\hline
Unfolded  &    $-0.010\pm0.043$       &     $0.006\pm0.031$     &      $0.015\pm0.025$     \\
Theory & $0.0026^{+0.0008}_{-0.0001}$ & $0.0066^{+0.0001}_{-0.0003}$ & $0.0202^{+0.0006}_{-0.0007}$ \\
\hline
\end{tabular}
}
\caption{\label{tab:results_diff}Measured charge asymmetry, \Ac, values for the electron and muon channels combined after
unfolding as a function of the \ttbar\ invariant mass, \mttbar\ (top), the \ttbar\ transverse momentum, \ptttbar (middle) and the \ttbar\ rapidity, \yttbar\ (bottom). SM predictions, as described in the text, are also reported. The quoted uncertainties include statistical and systematic components after the marginalisation.}
\end{center}
\end{table*}
\begin{table*}[tbp]
\begin{center}
{\footnotesize
\begin{tabular}{|l |c |c| c| c| c|}
\hline  
 &\multicolumn{5}{|c|}{$\mttbar{}$ ~[\GeV]} \\
\hline
Source of systematic uncertainty           &         $0$--$420$ & $420$--$500$      & $500$--$600$         & $600$--$750$      &   $>750$ \\ 
\hline
Lepton reconstruction/identification   	  	  & $<0.005$ & $<0.005$ & $<0.005$ & $<0.005$ & $<0.005$  \\ 
Lepton energy scale and resolution         	  & 0.017 & 0.014 & 0.013 & 0.007 & $<0.005$ \\
Jet energy scale and resolution            	  & 0.014 & 0.007 & 0.035 & 0.032 & 0.017 \\
Missing transverse momentum and pile--up modelling & 0.013 & 0.017 & 0.018 & 0.008 & 0.005 \\
Multi--jets background normalisation	& $<0.005$ & $<0.005$ & $<0.005$ & $<0.005$ & $<0.005$ \\
$b$--tagging/mis--tag efficiency		& $<0.005$ & $<0.005$ & $<0.005$ & $<0.005$ & $<0.005$ \\
Signal modelling			& $<0.005$ & $<0.005$ & $<0.005$ & $<0.005$ & $<0.005$ \\
Parton shower/hadronisation		& $<0.005$ & $<0.005$ & $<0.005$ & $<0.005$ & $<0.005$ \\
Monte Carlo sample size			& $<0.005$ & $<0.005$ & $<0.005$ & $<0.005$ & $<0.005$ \\
PDF					& $<0.005$ & $<0.005$ & $<0.005$ & $<0.005$ & $<0.005$ \\
$W$+jets normalisation and shape        & $<0.005$ & $<0.005$ & $<0.005$ & $<0.005$ & $<0.005$ \\
\hline
Statistical uncertainty			   & 0.054 	       & 	0.042	   &	    0.046      &       0.052       &         0.054     \\
\hline
\end{tabular}
}
\end{center}
\vspace*{2pt}
\begin{center}
{\footnotesize
\begin{tabular}{|l |c |c| c|}
\hline  
 &\multicolumn{3}{|c|}{$\ptttbar{}$ ~[\GeV]} \\
\hline
Source of systematic uncertainty           &    $0$--$25$           &        $25$--$60$         &  $>60$    \\ 
\hline
Lepton reconstruction/identification   	   	  & $<0.005$ & $<0.005$ & $<0.005$ \\ 
Lepton energy scale and resolution         	  & 0.011 & 0.013 & 0.006   \\
Jet energy scale and resolution            	  & 0.009 & 0.020 & 0.020   \\
Missing transverse momentum and pile--up modelling & 0.017 & 0.010 & $<0.005$ \\
Multi--jets background normalisation	& $<0.005$ & $<0.005$ & $<0.005$  \\
$b$--tagging/mis--tag efficiency		& $<0.005$ & $<0.005$ & $<0.005$  \\
Signal modelling			& $<0.005$ & $<0.005$ & $<0.005$  \\
Parton shower/hadronisation		& $<0.005$ & $<0.005$ & $<0.005$  \\
Monte Carlo sample size			& $<0.005$ & $<0.005$ & $<0.005$  \\
PDF					& $<0.005$ & $<0.005$ & $<0.005$  \\
$W$+jets normalisation and shape        & $<0.005$ & $<0.005$ & $<0.005$  \\
\hline
Statistical uncertainty			& 0.052 	       & 	0.057	   &	    0.034     \\
\hline
\end{tabular}
}
\end{center}
\vspace*{2pt}
\begin{center}
{\footnotesize
\begin{tabular}{|l |c |c| c|}
\hline  
 & \multicolumn{3}{|c|}{$\yttbar{}$}    \\
\hline
Source of systematic uncertainty           &    $0$--$0.3$          &        $0.3$--$0.7$     &  $>0.7$   \\ 
\hline
Lepton reconstruction/identification   	   	  & $<0.005$ & $<0.005$ & $<0.005$ \\ 
Lepton energy scale and resolution         	  & 0.022 & 0.014 & 0.008   \\
Jet energy scale and resolution            	  & 0.013 & 0.007 & $<0.005$   \\
Missing transverse momentum and pile--up modelling & $<0.005$ & 0.006 & $<0.005$ \\
Multi--jets background normalisation	& $<0.005$ & $<0.005$ & $<0.005$  \\
$b$--tagging/mis--tag efficiency		& $<0.005$ & $<0.005$ & $<0.005$  \\
Signal modelling			& $<0.005$ & $<0.005$ & $<0.005$  \\
Parton shower/hadronisation		& $<0.005$ & $<0.005$ & $<0.005$  \\
Monte Carlo sample size			& $<0.005$ & $<0.005$ & $<0.005$  \\
PDF					& $<0.005$ & $<0.005$ & $<0.005$  \\
$W$+jets normalisation and shape        & $<0.005$ & $<0.005$ & $<0.005$  \\
\hline
Statistical uncertainty			& 0.042 	       & 	0.030	   &	    0.025     \\
\hline
\end{tabular}
}
\caption{\label{table:Systematics_diff}Systematic uncertainties for the charge asymmetry, \Ac, measurement for the electron and muon channels combined after
unfolding as a function of the \ttbar\ invariant mass, \mttbar\ (top), the \ttbar\ transverse momentum, \ptttbar\ (middle) and the \ttbar\ rapidity, \yttbar\ (bottom). For variations resulting in asymmetric uncertainties, the average absolute deviation from the nominal value is reported. The values reported for 
each systematic uncertainty are the variation of the mean of posterior probability densities computed considering 1$\sigma$ variations.}
\end{center}
\end{table*}
\begin{table*}[tbp]
\begin{center}
{\small
\begin{tabular}{|l| c| c |c| c| c |}
\hline
 &\multicolumn{5}{|c|}{$\mttbar{}$ ~[\GeV]}    \\
\hline
$\rho_{i,j}$ & $0$--$420$ & $420$--$500$ & $500$--$600$ & $600$--$750$ & $>750$ \\
\hline
$0$--$420$      & $1$ & $-0.38$ & $0.13$  & $-0.05$ & $0.01$  \\
$420$--$500$    &     & $1$     & $-0.53$ & $0.17$  & $-0.03$ \\
$500$--$600$    &     &         & $1$     & $-0.54$ & $0.14$  \\
$600$--$750$    &     &         &         & $1$     & $-0.43$ \\
$>750$          &     &         &         &         & $1$     \\
\hline
\end{tabular}
}
\end{center}
\vspace*{2pt}
\begin{center}
{\small
\begin{tabular}{|l| c| c| c |}
\hline
  &\multicolumn{3}{|c|}{$\ptttbar{}$ ~[\GeV]}    \\
\hline
$\rho_{i,j}$      &     $0$--$25$           &        $25$--$60$         &  $>60$            \\
\hline
$0$--$25$       &    $1$    &    $-0.79$  &   $0.36$  \\
$25$--$60$      &           &    $1$      &   $-0.60$ \\
$>60$           &           &             &   $1$     \\
\hline
\end{tabular}
}
\end{center}
\vspace*{2pt}
\begin{center}
{\small
\begin{tabular}{|l| c| c| c |}
\hline
  &\multicolumn{3}{|c|}{$\yttbar~$}    \\
\hline
$\rho_{i,j}$      &  $0$--$0.3$   &  $0.3$--$0.7$  &  $>0.7$  \\
\hline
$0$--$0.3$    &   $1$   &   $-0.33$  &   $0.05$   \\
$0.3$--$0.7$  &         &   $1$      &   $-0.21$  \\
$>0.7$        &         &            &   $1$      \\
\hline
\end{tabular}
}
\caption{\label{table:varAC_diff}Correlation coefficients $\rho_{i,j}$ for the statistical uncertainties between the $i$--th and $j$--th bin of the differential \Ac\ measurement as 
a function of the \ttbar\ invariant mass, \mttbar\ (top), the transverse momentum, \ptttbar\ (middle) and the \ttbar\ rapidity, \yttbar\ (bottom).}
\end{center}
\end{table*}
\subsection{Measurements for $\beta_{z,\ttbar}>0.6$}
\label{beta_results}
An additional requirement on the $z$--component of the \ttbar--system velocity $\beta_{z,\ttbar}>0.6$ is applied, as explained in 
section~\ref{Introduction}, for the inclusive and the differential \DY{}distribution as a function of \mttbar. It has been verified that resolution effects on the reconstructed
$\beta_{z,\ttbar}$ did not introduce any bias in the measurement. Hence an unfolding of the $\beta_{z,\ttbar}$ distribution was found to be unnecessary. The inclusive asymmetry after this requirement is \resultinclbeta, as reported in the last row of table~\ref{tab:results}, to be compared with the SM prediction
$A_{\rm{C}}^{\rm{SM}}=0.020^{+0.006}_{-0.007}$~\cite{Bern:2012}. Table~\ref{Table_Systematics} (right column) shows the list of systematic uncertainties affecting the measurement before 
the marginalisation procedure. 

Figure~\ref{fig:DY_unf_diff} (bottom right plot) shows the differential \Ac\ measurement as a function of \mttbar, while table~\ref{tab:results_mtt_beta} shows 
the value of \Ac\ for the different bins, table~\ref{table:Systematics_mtt_beta} lists the systematic uncertainties affecting the measurement before the marginalisation and 
table~\ref{table:varAC_MttBeta_unf} shows the correlation coefficients among the different bins. These measurements do not deviate significantly from the SM expectations either.
\begin{table*}[tbp]
\begin{center}
{\small
\begin{tabular}{|l |c| c| c| c| c |}
\hline
 &\multicolumn{5}{|c|}{$\mttbar{}$ ~[\GeV] for $\beta_{z,\ttbar} > 0.6$}    \\
\hline
\Ac & $0$--$420$ & $420$--$500$ & $500$--$600$ & $600$--$750$ & $>750$ \\
\hline
Unfolded  & $0.054 \pm  0.079 $ & $0.008 \pm 0.072  $ & $-0.022 \pm 0.075 $ & $-0.019 \pm 0.102    $ & $0.205 \pm 0.135$ \\
Theory & $0.0145^{+0.0005}_{-0.0003}$ & $0.0213^{+0.0006}_{-0.0005}$ & $0.0240^{+0.0003}_{-0.0009}$ & $0.0280^{+0.0012}_{-0.0007}$ & $0.0607 \pm 0.0002$ \\
\hline
\end{tabular}
}
\caption{\label{tab:results_mtt_beta}Measured charge asymmetry, \Ac, values for the electron and muon channels combined after
unfolding as a function of the \ttbar\ invariant mass, \mttbar, for $\beta_{z,\ttbar} >0.6$. SM predictions, as described in the text, are also reported.
The quoted uncertainties include statistical and systematic components after the marginalisation.}
\end{center}
\end{table*}
\begin{table*}[tbp]
\begin{center}
{\footnotesize
\begin{tabular}{|l |c |c| c| c| c|}
\hline  
 &\multicolumn{5}{|c|}{$\mttbar{}$ ~[\GeV] for $\beta_{z,\ttbar} > 0.6$}    \\
\hline
Source of systematic uncertainty           &         $0$--$420$ & $420$--$500$      & $500$--$600$         & $600$--$750$      &   $>750$ \\ 
\hline
Lepton reconstruction/identification   	  	  & $<0.005$ & $<0.005$ & $<0.005$ & $<0.005$ & $<0.005$  \\ 
Lepton energy scale and resolution         	  & 0.021 & 0.033 & 0.039 & 0.024 & 0.015 \\
Jet energy scale and resolution            	  & 0.014 & 0.026 & 0.061 & 0.095 & 0.111 \\
Missing transverse momentum and pile--up modelling & 0.019 & 0.030 & 0.032 & 0.019 & 0.011 \\
Multi--jets background normalisation	& 0.007 & $<0.005$ & $<0.005$ & $<0.005$ & 0.017 \\
$b$--tagging/mis--tag efficiency		& $<0.005$ & $<0.005$ & $<0.005$ & $<0.005$ & $<0.005$ \\
Signal modelling			& $<0.005$ & $<0.005$ & $<0.005$ & $<0.005$ & $<0.005$ \\
Parton shower/hadronisation		& $<0.005$ & $<0.005$ & $<0.005$ & $<0.005$ & $<0.005$ \\
Monte Carlo sample size			& $<0.005$ & $<0.005$ & $<0.005$ & $<0.005$ & $<0.005$ \\
PDF					& $<0.005$ & $<0.005$ & $<0.005$ & $<0.005$ & $<0.005$ \\
$W$+jets normalisation and shape        & $<0.005$ & $<0.005$ & $<0.005$ & $<0.005$ & 0.010 \\
\hline
Statistical uncertainty			   & 0.078 	       & 	0.070	   &	    0.074      &       0.098       &         0.131     \\
\hline
\end{tabular}
}
\caption{\label{table:Systematics_mtt_beta}Systematic uncertainties for the charge asymmetry, \Ac, measurement for the electron and muon channels combined after
unfolding as a function of the \ttbar\ invariant mass, \mttbar, for $\beta_{z,\ttbar} > 0.6$. For variations resulting in asymmetric uncertainties, the average absolute deviation 
from the nominal value is reported. The values reported for each systematic uncertainty are the variation of the mean of posterior probability densities computed considering 1$\sigma$ variations.}
\end{center}
\end{table*}
\begin{table*}[tbp]
\begin{center}
{\small
\begin{tabular}{|l| c| c| c| c |c| }
\hline
 &\multicolumn{5}{|c|}{$\mttbar{}$ ~[\GeV] for $\beta_{z,\ttbar}> 0.6$}    \\
\hline
$\rho_{i,j}$ & $0$--$420$ & $420$--$500$ & $500$--$600$ & $600$--$750$ & $>750$ \\
\hline
$0$--$420$      & $1$ & $-0.36$ & $0.08$  & $-0.01$ & $0.01$  \\
$420$--$500$    &     & $1$     & $-0.57$ & $0.19$  & $-0.04$ \\
$500$--$600$    &     &         & $1$     & $-0.59$ & $0.16$  \\
$600$--$750$    &     &         &         & $1$     & $-0.50$ \\
$>750$          &     &         &         &         & $1$     \\
\hline
\end{tabular}
}
\caption{\label{table:varAC_MttBeta_unf}Correlation coefficients $\rho_{i,j}$ for the statistical uncertainties between the $i$--th and $j$--th bin of the differential \Ac\ measurement as a function of the
\ttbar\ invariant mass, \mttbar, for $\beta_{z,\ttbar} > 0.6$.}
\end{center}
\end{table*}

\subsection{Interpretation}
\label{Interpretation}
\begin{figure}[htp]
\begin{center}
\includegraphics[width=0.45\textwidth]{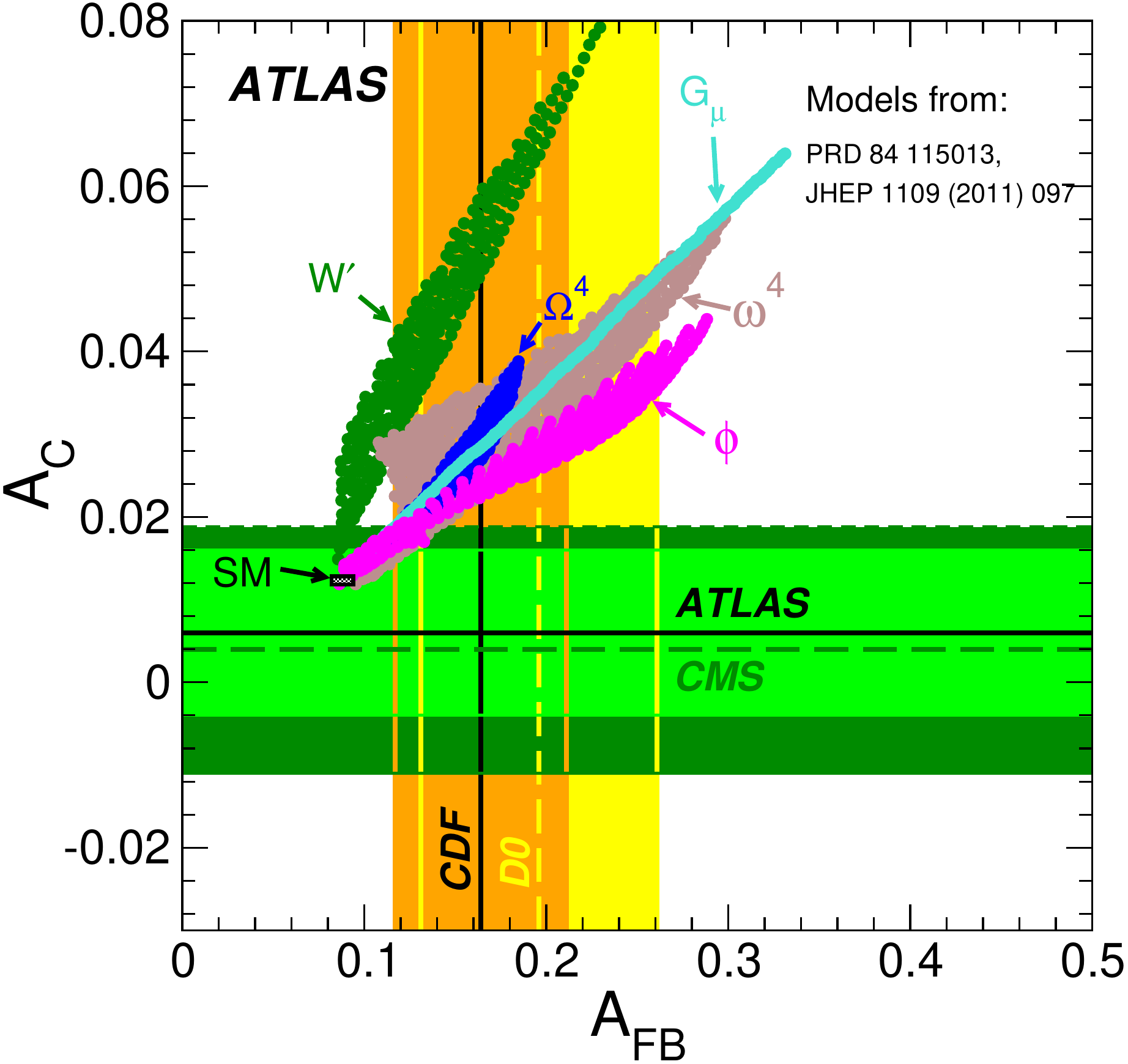}
\includegraphics[width=0.45\textwidth]{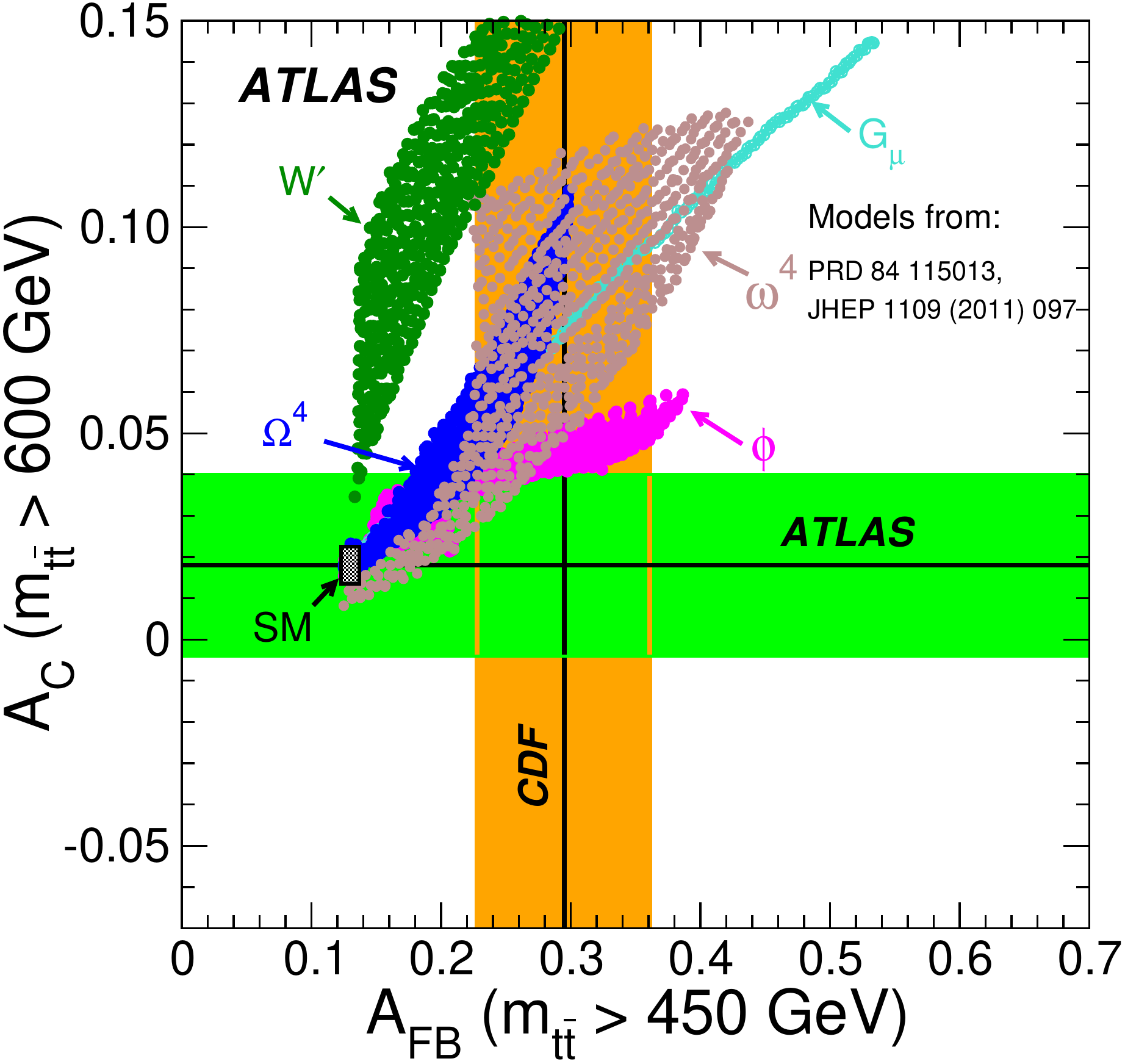}
\caption{Measured forward--backward asymmetries $A_{\rm{FB}}$ at Tevatron and charge asymmetries \Ac\ at LHC, 
compared with the SM predictions (black box) as well as predictions incorporating various potential new physics contributions (as described in the figure)~\cite{AguilarSaavedra:2011hz,AguilarSaavedra:2011ug}.
In both plots, where present, the horizontal bands and lines correspond to the ATLAS (light green) and CMS (dark green) measurements, while the vertical ones correspond to
the CDF (orange) and D0 (yellow) measurements.
The inclusive \Ac\ measurements are reported in the left plot. In the right plot a comparison is reported between the $A_{\rm{FB}}$ measurement by CDF for $\mttbar > 450$ \GeV\ and the 
\Ac\ measurement for $\mttbar > 600$ \GeV.}
\label{fig:AvsA}
\end{center}
\end{figure}

Figure~\ref{fig:AvsA} shows the inclusive \Ac\ measurements with and without the additional requirement on the invariant mass of the \ttbar--system 
$\mttbar > 600$ \GeV\ described in section~\ref{Results}. In the left plot, the \Ac\ measurement without the $\mttbar > 600$ \GeV\ requirement is compared with the corresponding measurement from 
CMS~\cite{CMS_ljets2} (horizontal lines) and with the \ttbar\ forward--backward asymmetry $A_{\rm{FB}}$ measurements made at the Tevatron by CDF, 
$A_{\rm{FB}} = 0.164 \pm 0.045$~\cite{CDF3}, and D0, $A_{\rm{FB}} = 0.196 \pm 0.065$~\cite{D02} (vertical lines). In the right plot, the \Ac\ measurement with the requirement of 
$\mttbar > 600$ \GeV, is compared with the $A_{\rm{FB}}$ measurement, with the requirement of $\mttbar > 450$ \GeV, performed by the CDF experiment at the Tevatron~\cite{CDF3}.

Predictions given by several new physics models introduced to explain the larger than expected $A_{\rm{FB}}$ values measured at the Tevatron are also displayed. Details of these models can be found in 
refs.~\cite{ATLAS_ljets,AguilarSaavedra:2011hz,AguilarSaavedra:2011ug}.
For each model, the predictions for $A_{\rm{FB}}$ and \Ac\ are derived using the
{\sc PROTOS} generator~\cite{AguilarSaavedra:2008gt} with the constraints described in ref.~\cite{ATLAS_ljets}.
The ranges of predicted values for $A_{\rm{FB}}$ and \Ac\ for a given new physics model are also shown. The new physics contributions are computed using the tree--level SM 
amplitude plus the one(s) from the new particle(s), to account for the interference between the two contributions. Some of these new physics models seem to be disfavoured by the current
measurements.
\clearpage

%\section{Comparison of LHC and Tevatron Results}
%\label{Interpretation}
\section{Conclusion}
\label{Conclusion}
This paper has presented a measurement of the \ttbar\ production charge asymmetry measurement in \ttbar--events with a single lepton (electron or muon), at least four jets, of which at 
least one is tagged as a $b$--jet, and large missing transverse momentum, using an integrated luminosity of \lumi\ recorded by the ATLAS experiment in $pp$ collisions at a 
centre--of--mass energy of $\sqrt{s}$ = 7 TeV at the LHC. 
The inclusive \ttbar\ production charge asymmetry \Ac\ and its differential distributions, as a function of \mttbar, \ptttbar\ and \yttbar, have been unfolded to parton--level. 
The measured inclusive \ttbar\ production charge asymmetry is \resultincl, to be compared with the SM prediction $A_{\rm{C}}^{\text{SM}}=0.0123\pm0.0005$. All measurements presented are 
statistically limited and are found to be compatible with the SM prediction within the uncertainties.

%\section*{Acknowledgements}
%\label{Acknowledgements}
%\section{Tables}
\acknowledgments
% Acknowledgements for papers with collision data
% Version 14-Nov-2013
%\section{Acknowledgements}
% Standard acknowledgements start here
%----------------------------------------------
We thank CERN for the very successful operation of the LHC, as well as the
support staff from our institutions without whom ATLAS could not be
operated efficiently.

We acknowledge the support of ANPCyT, Argentina; YerPhI, Armenia; ARC,
Australia; BMWF and FWF, Austria; ANAS, Azerbaijan; SSTC, Belarus; CNPq and FAPESP,
Brazil; NSERC, NRC and CFI, Canada; CERN; CONICYT, Chile; CAS, MOST and NSFC,
China; COLCIENCIAS, Colombia; MSMT CR, MPO CR and VSC CR, Czech Republic;
DNRF, DNSRC and Lundbeck Foundation, Denmark; EPLANET, ERC and NSRF, European Union;
IN2P3-CNRS, CEA-DSM/IRFU, France; GNSF, Georgia; BMBF, DFG, HGF, MPG and AvH
Foundation, Germany; GSRT and NSRF, Greece; ISF, MINERVA, GIF, DIP and Benoziyo Center,
Israel; INFN, Italy; MEXT and JSPS, Japan; CNRST, Morocco; FOM and NWO,
Netherlands; BRF and RCN, Norway; MNiSW and NCN, Poland; GRICES and FCT, Portugal; MNE/IFA, Romania; MES of Russia and ROSATOM, Russian Federation; JINR; MSTD,
Serbia; MSSR, Slovakia; ARRS and MIZ\v{S}, Slovenia; DST/NRF, South Africa;
MINECO, Spain; SRC and Wallenberg Foundation, Sweden; SER, SNSF and Cantons of
Bern and Geneva, Switzerland; NSC, Taiwan; TAEK, Turkey; STFC, the Royal
Society and Leverhulme Trust, United Kingdom; DOE and NSF, United States of
America.

The crucial computing support from all WLCG partners is acknowledged
gratefully, in particular from CERN and the ATLAS Tier-1 facilities at
TRIUMF (Canada), NDGF (Denmark, Norway, Sweden), CC-IN2P3 (France),
KIT/GridKA (Germany), INFN-CNAF (Italy), NL-T1 (Netherlands), PIC (Spain),
ASGC (Taiwan), RAL (UK) and BNL (USA) and in the Tier-2 facilities
worldwide.
%--------------------------------------------

\bibliographystyle{atlasnote}
\bibliography{TopChargeAsym_CONF,PHYSICS,ATLAS}

\providecommand{\href}[2]{#2}\begingroup\raggedright\begin{thebibliography}{10}

\bibitem{Jung:2011zv}
S.~Jung, A.~Pierce, and J.~D. Wells, {\em {Top quark asymmetry from a
  non-Abelian horizontal symmetry}\/},
  \href{http://dx.doi.org/10.1103/PhysRevD.83.114039}{Phys. Rev. { D 83} (2011)
   114039},
\href{http://arxiv.org/abs/1103.4835}{{\tt arXiv:1103.4835 [hep-ph]}}.
%%CITATION = 1103.4835;%%.

\bibitem{AXI}
O.~Antu{\~{n}}ano, J.~H. K{\"{u}}hn, and G.~Rodrigo, {\em Top quarks,
  axigluons, and charge asymmetries at hadron colliders\/},  Phys. Rev. { D 77}
  (2008)  014003, \href{http://arxiv.org/abs/0709.1652}{{\tt arXiv:0709.1652
  [hep-ph]}}.

\bibitem{Djouadi:2009nb}
A.~Djouadi, G.~Moreau, F.~Richard, and R.~K. Singh, {\em {Forward-backward
  asymmetry of top quark production at the Tevatron in warped extra dimensional
  models}\/},  \href{http://dx.doi.org/10.1103/PhysRevD.82.071702}{Phys. Rev. {
  D 82} (2010)  071702(R)},
\href{http://arxiv.org/abs/0906.0604}{{\tt arXiv:0906.0604 [hep-ph]}}.
%%CITATION = 0906.0604;%%.

\bibitem{KK}
P.~Ferrario and G.~Rodrigo, {\em Massive color-octet bosons and the charge
  asymmetries of top quarks at hadron colliders\/},  Phys. Rev. { D 78} (2008)
  094018, \href{http://arxiv.org/abs/0809.3354}{{\tt arXiv:0809.3354
  [hep-ph]}}.

\bibitem{Jung:2009jz}
S.~Jung, H.~Murayama, A.~Pierce, and J.~D. Wells, {\em {Top quark
  forward-backward asymmetry from new t-channel physics}\/},
  \href{http://dx.doi.org/10.1103/PhysRevD.81.015004}{Phys. Rev. { D 81} (2010)
   015004}, \href{http://arxiv.org/abs/0907.4112}{{\tt arXiv:0907.4112
  [hep-ph]}}.

\bibitem{Shu:2009xf}
J.~Shu, T.~M.~P. Tait, and K.~Wang, {\em {Explorations of the top quark
  forward-backward asymmetry at the Tevatron}\/},
  \href{http://dx.doi.org/10.1103/PhysRevD.81.034012}{Phys. Rev. { D 81} (2010)
   034012}, \href{http://arxiv.org/abs/0911.3237}{{\tt arXiv:0911.3237
  [hep-ph]}}.

\bibitem{JA:2011}
J.~A. Aguilar-Saavedra and M.~P\'erez-Victoria, {\em {Probing the Tevatron
  $t\bar{t}$ asymmetry at LHC}\/},
  \href{http://dx.doi.org/10.1007/JHEP05(2011)034}{JHEP { 1105} (2011)  034}.

\bibitem{AguilarSaavedra:2011hz}
J.~A. Aguilar-Saavedra and M.~P\'erez-Victoria, {\em {Asymmetries in $t\bar{t}$
  production: LHC versus Tevatron}\/},
  \href{http://dx.doi.org/10.1103/PhysRevD.84.115013}{Phys. Rev. { D 84} (2011)
   115013},
\href{http://arxiv.org/abs/1105.4606}{{\tt arXiv:1105.4606 [hep-ph]}}.
%%CITATION = ARXIV:1105.4606;%%.

\bibitem{Dorsner:2009mq}
I.~Dor\v{s}ner, S.~Fajfer, J.~F. Kamenik, and N.~Ko\v{s}nik, {\em {Light
  colored scalars from grand unification and the forward-backward asymmetry in
  $t\bar{t}$ production}\/},
  \href{http://dx.doi.org/10.1103/PhysRevD.81.055009}{Phys. Rev. { D 81} (2010)
   055009},
\href{http://arxiv.org/abs/0912.0972}{{\tt arXiv:0912.0972 [hep-ph]}}.
%%CITATION = 0912.0972;%%.

\bibitem{Grinstein:2011yv}
B.~Grinstein, A.~L. Kagan, M.~Trott, and J.~Zupan, {\em {Forward-backward
  asymmetry in $t\bar{t}$ production from flavor symmetries}\/},
  \href{http://dx.doi.org/10.1103/PhysRevLett.107.012002}{Phys. Rev. Lett. {
  107} (2011)  012002},
\href{http://arxiv.org/abs/1102.3374}{{\tt arXiv:1102.3374 [hep-ph]}}.
%%CITATION = 1102.3374;%%.

\bibitem{Ligeti:2011vt}
Z.~Ligeti, G.~M. Tavares, and M.~Schmaltz, {\em {Explaining the $t\bar{t}$
  forward-backward asymmetry without dijet or flavor anomalies}\/},
  \href{http://dx.doi.org/10.1007/JHEP06(2011)109}{JHEP { 1106} (2011)  109},
\href{http://arxiv.org/abs/1103.2757}{{\tt arXiv:1103.2757 [hep-ph]}}.
%%CITATION = 1103.2757;%%.

\bibitem{Ferrario:2009bz}
P.~Ferrario and G.~Rodrigo, {\em {Constraining heavy colored resonances from
  top-antitop quark events}\/},
  \href{http://dx.doi.org/10.1103/PhysRevD.80.051701}{Phys. Rev. { D 80} (2009)
   051701(R)},
\href{http://arxiv.org/abs/0906.5541}{{\tt arXiv:0906.5541 [hep-ph]}}.
%%CITATION = 0906.5541;%%.

\bibitem{Frampton:2009rk}
P.~H. Frampton, J.~Shu, and K.~Wang, {\em {Axigluon as possible explanation for
  $p\bar{p} \to t\bar{t}$ forward-backward asymmetry}\/},
  \href{http://dx.doi.org/10.1016/j.physletb.2009.12.043}{Phys. Lett. { B 683}
  (2010)  294--297},
\href{http://arxiv.org/abs/0911.2955}{{\tt arXiv:0911.2955 [hep-ph]}}.
%%CITATION = 0911.2955;%%.

\bibitem{AguilarSaavedra:2011ci}
J.~A. Aguilar-Saavedra and M.~P\'erez-Victoria, {\em {Shaping the top
  asymmetry}\/},  \href{http://dx.doi.org/10.1016/j.physletb.2011.10.004}{Phys.
  Lett. { B 705} (2011)  228},
\href{http://arxiv.org/abs/1107.2120}{{\tt arXiv:1107.2120 [hep-ph]}}.
%%CITATION = ARXIV:1107.2120;%%.

\bibitem{Kuhn:1998kw}
J.~H. K{\"{u}}hn and G.~Rodrigo, {\em {Charge asymmetry of heavy quarks at
  hadron colliders}\/},
  \href{http://dx.doi.org/10.1103/PhysRevD.59.054017}{Phys. Rev. { D 59} (1999)
   054017},
\href{http://arxiv.org/abs/hep-ph/9807420}{{\tt arXiv:hep-ph/9807420}}.
%%CITATION = HEP-PH/9807420;%%.

\bibitem{Kuhn:1998jr}
J.~H. K{\"{u}}hn and G.~Rodrigo, {\em {Charge asymmetry in hadroproduction of
  heavy quarks}\/},  \href{http://dx.doi.org/10.1103/PhysRevLett.81.49}{Phys.
  Rev. Lett. { 81} (1998)  49},
\href{http://arxiv.org/abs/hep-ph/9802268}{{\tt arXiv:hep-ph/9802268}}.
%%CITATION = HEP-PH/9802268;%%.

\bibitem{Bernreuther:2010ny}
W.~Bernreuther and Z.-G. Si, {\em {Distributions and correlations for top quark
  pair production and decay at the Tevatron and LHC}\/},
  \href{http://dx.doi.org/10.1016/j.nuclphysb.2010.05.001}{Nucl. Phys. { B 837}
  (2010)  90},
\href{http://arxiv.org/abs/1003.3926}{{\tt arXiv:1003.3926 [hep-ph]}}.
%%CITATION = 1003.3926;%%.

\bibitem{Ahrens:2011uf}
V.~Ahrens et al., {\em {Top-pair forward-backward asymmetry beyond
  next-to-leading order}\/},
  \href{http://dx.doi.org/10.1103/PhysRevD.84.074004}{Phys. Rev. { D 84} (2011)
   074004},
\href{http://arxiv.org/abs/1106.6051}{{\tt arXiv:1106.6051 [hep-ph]}}.
%%CITATION = 1106.6051;%%.

\bibitem{Hollik:2011ps}
W.~Hollik and D.~Pagani, {\em {Electroweak contribution to the top quark
  forward-backward asymmetry at the Tevatron}\/},
  \href{http://dx.doi.org/10.1103/PhysRevD.84.093003}{Phys. Rev. { D 84} (2011)
   093003},
\href{http://arxiv.org/abs/1107.2606}{{\tt arXiv:1107.2606 [hep-ph]}}.
%%CITATION = ARXIV:1107.2606;%%.

\bibitem{Kuhn:2011ri}
J.~H. K{\"{u}}hn and G.~Rodrigo, {\em {Charge asymmetries of top quarks at
  hadron colliders revisited}\/},  JHEP { 1201} (2012)  063,
  \href{http://arxiv.org/abs/1109.6830}{{\tt arXiv:1109.6830 [hep-ph]}}.

\bibitem{Bern:2012}
W.~Bernreuther and Z.-G. Si, {\em {Top quark and leptonic charge asymmetries
  for the Tevatron and LHC}\/},  Phys. Rev. { D 86} (2012)  034026,
  \href{http://arxiv.org/abs/1205.6580}{{\tt arXiv:1205.6580 [hep-ph]}}.

\bibitem{CDF1}
{CDF} Collaboration, T.~Aaltonen et al., {\em Forward-backward asymmetry in
  top-quark production in $p\bar{p}$ collisions at $\sqrt{s} =$ 1.96
  {T}e{V}\/},  Phys. Rev. Lett. { 101} (2008)  202001,
  \href{http://arxiv.org/abs/0806.2472}{{\tt arXiv:0806.2472 [hep-ex]}}.

\bibitem{CDF2}
{CDF} Collaboration, T.~Aaltonen et al., {\em Evidence for a mass dependent
  forward-backward asymmetry in top quark pair production\/},  Phys. Rev. { D
  83} (2011)  112003, \href{http://arxiv.org/abs/1101.0034}{{\tt
  arXiv:1101.0034 [hep-ex]}}.

\bibitem{CDF3}
{CDF} Collaboration, T.~Aaltonen et al., {\em Measurement of the top quark
  forward-backward production asymmetry and its dependence on event kinematic
  properties\/},  Phys. Rev. { D 87} (2013)  092002,
  \href{http://arxiv.org/abs/1211.1003}{{\tt arXiv:1211.1003 [hep-ex]}}.

\bibitem{D01}
{D0} Collaboration, V.~M. Abazov et al., {\em Measurement of the
  forward-backward charge asymmetry in top-quark pair production\/},  Phys.
  Rev. Lett. { 100} (2008)  142002, \href{http://arxiv.org/abs/0712.0851}{{\tt
  arXiv:0712.0851 [hep-ex]}}.

\bibitem{D02}
{D0} Collaboration, V.~M. Abazov et al., {\em {Forward-backward asymmetry in
  top quark-antiquark production}\/},
  \href{http://dx.doi.org/10.1103/PhysRevD.84.112005}{Phys. Rev. { D 84} (2011)
   112005},
\href{http://arxiv.org/abs/1107.4995}{{\tt arXiv:1107.4995 [hep-ex]}}.
%%CITATION = ARXIV:1107.4995;%%.

\bibitem{Diener:2009ee}
R.~Diener, S.~Godfrey, and T.~A.~W. Martin, {\em {Using final state
  pseudorapidities to improve s-channel resonance observables at the LHC}\/},
  \href{http://dx.doi.org/10.1103/PhysRevD.80.075014}{Phys. Rev. { D 80} (2009)
   075014},
\href{http://arxiv.org/abs/0909.2022}{{\tt arXiv:0909.2022 [hep-ph]}}.
%%CITATION = 0909.2022;%%.

\bibitem{CMS_ljets}
{CMS} Collaboration, {\em {Measurement of the charge asymmetry in top-quark
  pair production in proton-proton collisions at $\sqrt{s} = 7$ TeV}\/},
  \href{http://dx.doi.org/10.1016/j.physletb.2012.01.078}{Phys. Lett. B { 709}
  (2012)  28},
\href{http://arxiv.org/abs/1112.5100}{{\tt arXiv:1112.5100 [hep-ex]}}.
%%CITATION = ARXIV:1112.5100;%%.

\bibitem{CMS_ljets2}
{CMS} Collaboration, {\em {Inclusive and differential measurements of the
  \ttbar\ charge asymmetry in proton-proton collisions at $\sqrt{s} = 7$
  TeV}\/},  \href{http://dx.doi.org/10.1016/j.physletb.2012.09.028}{Phys. Lett.
  B { 717} (2012)  129}, \href{http://arxiv.org/abs/1207.0065}{{\tt
  arXiv:1207.0065 [hep-ex]}}.

\bibitem{ATLAS_ljets}
{ATLAS} Collaboration, {\em {Measurement of the charge asymmetry in top quark
  pair production in pp collisions at $\sqrt{s} = 7$ TeV using the ATLAS
  detector}\/},  \href{http://dx.doi.org/10.1140/epjc/s10052-012-2039-5}{Eur.
  Phys. J. { C 72} (2012)  2039},
\href{http://arxiv.org/abs/1203.4211}{{\tt arXiv:1203.4211 [hep-ex]}}.
%%CITATION = ARXIV:1203.4211;%%.

\bibitem{AguilarSaavedra:2012prl}
J.~A. Aguilar-Saavedra and A.~Juste, {\em Collider independent $t\bar{t}$
  forward-backward asymmetry\/},  Phys. Rev. Lett. { 109} (2012)  211804,
  \href{http://arxiv.org/abs/1205.1898}{{\tt arXiv:1205.1898 [hep-ph]}}.

\bibitem{beta_cut}
J.~A. Aguilar-Saavedra, A.~Juste, and F.~Rubbo, {\em {Boosting the \ttbar\
  charge asymmetry}\/},
  \href{http://dx.doi.org/10.1016/physletb.2011.12.007}{Phys. Lett. B { 707}
  (2012)  92}, \href{http://arxiv.org/abs/1109.3710}{{\tt arXiv:1109.3710
  [hep-ph]}}.

\bibitem{Aad:2008zzm}
{ATLAS} Collaboration, {\em {The ATLAS Experiment at the CERN Large Hadron
  Collider}\/},
\href{http://dx.doi.org/10.1088/1748-0221/3/08/S08003}{JINST { 3} (2008)
  S08003}.
%%CITATION = JINST,3,S08003;%%.

\bibitem{lumi2011}
{ATLAS} Collaboration, {\em Improved luminosity determination in $pp$
  collisions at $\sqrt{s} = 7$ TeV using the ATLAS detector at the LHC\/},
  \href{http://dx.doi.org/10.1140/epjc/s10052-013-2518-3}{Eur. Phys. J. { C 73}
  (2013)  2518},
\href{http://arxiv.org/abs/1302.4393}{{\tt arXiv:1302.4393 [hep-ex]}}.
%%CITATION = ARXIV:1302.4393;%%.

\bibitem{MAN-0301-2}
{M.L. Mangano} et al., {\em ALPGEN, a generator for hard multiparton processes
  in hadronic collisions\/},  JHEP { 0307} (2003)  001,
  \href{http://arxiv.org/abs/hep-ph/0206293}{{\tt arXiv:hep-ph/0206293}}.

\bibitem{cteq61}
{J.~Pumplin et al.}, {\em {New generation of parton distributions with
  uncertainties from global QCD analysis}\/},
  \href{http://dx.doi.org/10.1088/1126-6708/2002/07/012}{JHEP { 0207} (2002)
  012}.

\bibitem{COR-0001-2}
{G. Corcella} et al., {\em HERWIG 6: An event generator for hadron emission
  reactions with interfering gluons (including supersymmetric processes)\/},
  JHEP { 0101} (2001)  010, \href{http://arxiv.org/abs/hep-ph/0011363}{{\tt
  arXiv:hep-ph/0011363}}.

\bibitem{JButterworth:1996zw}
J.~M. Butterworth, J.~R. Forshaw, and M.~H. Seymour, {\em {Multiparton
  interactions in photoproduction at HERA}\/},
  \href{http://dx.doi.org/10.1007/s002880050286}{Z. Phys. { C 72} (1996)  637},
\href{http://arxiv.org/abs/hep-ph/9601371}{{\tt arXiv:hep-ph/9601371}}.
%%CITATION = HEP-PH/9601371;%%.

\bibitem{Jimmy_tuning}
{ATLAS} Collaboration, {\em New ATLAS event generator tunes to 2010 data\/},
  ATL-PHYS-PUB-2011-008,
  \href{http://cds.cern.ch/record/1345343}{http://cds.cern.ch/record/1345343}
  .

\bibitem{Cacciari:2012}
{M.~Cacciari et al.}, {\em Top-pair production at hadron colliders with
  next-to-next-to-leading logarithmic soft-gluon resummation\/},  Phys. Lett. {
  B 710} (2012)  612, \href{http://arxiv.org/abs/1111.5869}{{\tt
  arXiv:1111.5869 [hep-ph]}}.

\bibitem{Barn:2012}
{P.~B\"arnreuther et al.}, {\em Percent level precision physics at the
  Tevatron: first genuine NNLO QCD corrections to $q\bar{q}\rightarrow
  t\bar{t}$\/},  Phys. Rev. Lett. { 109} (2012)  132001,
  \href{http://arxiv.org/abs/1204.5201}{{\tt arXiv:1204.5201 [hep-ph]}}.

\bibitem{Mitov1}
M.~Czakon and A.~Mitov, {\em NNLO corrections to top-pair production at hadron
  colliders: the all-fermionic scattering channels\/},  JHEP { 1212} (2012)
  054, \href{http://arxiv.org/abs/1207.0236}{{\tt arXiv:1207.0236 [hep-ph]}}.

\bibitem{Mitov2}
M.~Czakon and A.~Mitov, {\em NNLO corrections to top pair production at hadron
  colliders: the quark-gluon reaction\/},  JHEP { 1301} (2013)  080,
  \href{http://arxiv.org/abs/1210.6832}{{\tt arXiv:1210.6832}}.

\bibitem{Mitov3}
P.~F. M.~Czakon and A.~Mitov, {\em The total top quark pair production
  cross-section at hadron colliders through $\mathcal{O}(\alpha_S^4)$\/},
  \href{http://arxiv.org/abs/1303.6254}{{\tt arXiv:1303.6254 [hep-ph]}}.

\bibitem{Mitov4}
M.~Czakon and A.~Mitov, {\em Top++: a program for the calculation of the
  top-pair cross-section at hadron colliders\/},
  \href{http://arxiv.org/abs/1112.5675}{{\tt arXiv:1112.5675 [hep-ph]}}.

\bibitem{pdf4lhc}
{M.~Botje et al.}, {\em {The PDF4LHC Working Group Interim Recommendations}\/},
   \href{http://arxiv.org/abs/1101.0538}{{\tt arXiv:1101.0538 [hep-ph]}}.

\bibitem{AcerMC35}
B.~P. Kersevan and E.~Richter-Was, {\em The Monte Carlo event generator AcerMC
  versions 2.0 to 3.8 with interfaces to PYTHIA 6.4, HERWIG 6.5 and ARIADNE
  4.1\/},  Comput. Phys. Commun. { 184} (2013)  919,
  \href{http://arxiv.org/abs/hep-ph/0405247}{{\tt arXiv:hep-ph/0405247
  [hep-ph]}}.

\bibitem{AGO-0301-2}
{S. Agostinelli} et al., {\em Geant4 -- A Simulation Toolkit\/},  Nucl.
  Instrum. Meth. { A 506} (2003)  250.

\bibitem{2010EPJC...70..823A}
{ATLAS} Collaboration, {\em {The ATLAS Simulation Infrastructure}\/},  {Eur.
  Phys. J.} { C 70} (2010)  823, \href{http://arxiv.org/abs/1005.4568}{{\tt
  arXiv:1005.4568 [physics.ins-det]}}.

\bibitem{Cacciari:2008gp}
M.~Cacciari, G.~P. Salam, and G.~Soyez, {\em {The anti-$k_t$ jet clustering
  algorithm}\/},  \href{http://dx.doi.org/10.1088/1126-6708/2008/04/063}{JHEP {
  0804} (2008)  063}, \href{http://arxiv.org/abs/0802.1189}{{\tt
  arXiv:0802.1189 [hep-ph]}}.

\bibitem{ATL-CONF-2013-004}
{ATLAS} Collaboration, {\em {Jet energy scale and its systematic uncertainty in
  proton-proton collisions at $\sqrt{s}=7$ TeV with ATLAS 2011 data}\/},
  ATLAS-CONF-2013-004,
  \href{http://cds.cern.ch/record/1509552}{http://cds.cern.ch/record/1509552}
  .

\bibitem{JetFitter}
{ATLAS} Collaboration, {\em Commissioning of the ATLAS high-performance
  $b$-tagging algorithms in the 7 TeV collision data\/},  ATLAS-CONF-2011-102,
  \href{http://cds.cern.ch/record/1369219}{http://cds.cern.ch/record/1369219}
  .

\bibitem{SV0Tagger}
{ATLAS} Collaboration, {\em Calibrating the b-Tag Efficiency and Mistag Rate in
  35 pb$^{-1}$ of Data with the ATLAS Detector\/},  ATLAS-CONF-2011-089,
  \href{http://cds.cern.ch/record/1356198}{http://cds.cern.ch/record/1356198}
  .

\bibitem{Aad:2010ey}
{ATLAS} Collaboration, {\em {Measurement of the top quark-pair production cross
  section with ATLAS in pp collisions at $\sqrt{s}=7$ TeV}\/},
  \href{http://dx.doi.org/10.1140/epjc/s10052-011-1577-6}{Eur. Phys. J. { C 71}
  (2011)  1577}, \href{http://arxiv.org/abs/1012.1792}{{\tt arXiv:1012.1792
  [hep-ex]}}.

\bibitem{Fbu2012arXiv1201.4612C}
G.~{Choudalakis}, {\em {Fully Bayesian Unfolding}\/},
  \href{http://arxiv.org/abs/1201.4612}{{\tt arXiv:1201.4612
  [physics.data-an]}}.

\bibitem{D'Agostini:1994zf}
G.~D'Agostini, {\em {A multidimensional unfolding method based on Bayes'
  theorem}\/},
\href{http://dx.doi.org/10.1016/0168-9002(95)00274-X}{Nucl. Instrum. Meth. { A
  362} (1995)  487}.
%%CITATION = NUIMA,A362,487;%%.

\bibitem{ATLAS-CONF-2010-054}
{ATLAS} Collaboration, {\em Jet energy resolution in proton-proton collision at
  $\sqrt{s}$ = 7 TeV recorded in 2010 with the ATLAS detector\/},
  \href{http://dx.doi.org/10.1140/epjc/s10052-013-2306-0}{Eur. Phys. J. { C 73}
  (2013)  2306}, \href{http://arxiv.org/abs/1210.6210}{{\tt arXiv:1210.6210
  [hep-ex]}}.

\bibitem{nason:2004rx}
P.~Nason, {\em {A new method for combining NLO QCD with shower Monte Carlo
  algorithms}\/},  \href{http://dx.doi.org/10.1088/1126-6708/2004/11/040}{JHEP
  { 0411} (2004)  040},
\href{http://arxiv.org/abs/hep-ph/0409146}{{\tt arXiv:hep-ph/0409146}}.
%%CITATION = HEP-PH/0409146;%%.

\bibitem{Frixione:2007vw}
S.~Frixione, P.~Nason, and C.~Oleari, {\em {Matching NLO QCD computations with
  parton shower simulations: the POWHEG method}\/},  JHEP { 0711} (2007)  070,
\href{http://arxiv.org/abs/0709.2092}{{\tt arXiv:0709.2092 [hep-ph]}}.
%%CITATION = 0709.2092;%%.

\bibitem{AguilarSaavedra:2011ug}
J.~Aguilar-Saavedra and M.~Perez-Victoria, {\em {Simple models for the top
  asymmetry: Constraints and predictions}\/},  JHEP { 1109} (2011)  097,
\href{http://arxiv.org/abs/1107.0841}{{\tt arXiv:1107.0841 [hep-ph]}}.
%%CITATION = ARXIV:1107.0841;%%.

\bibitem{AguilarSaavedra:2008gt}
J.~Aguilar-Saavedra, {\em {Single top quark production at LHC with anomalous
  Wtb couplings}\/},
  \href{http://dx.doi.org/10.1016/j.nuclphysb.2008.06.013}{Nucl. Phys. { B 804}
  (2008)  160--192}, \href{http://arxiv.org/abs/0803.3810}{{\tt arXiv:0803.3810
  [hep-ph]}}.

\end{thebibliography}\endgroup
\onecolumn
\clearpage 
% ATLAS Collaboration author list
% Data extracted on 12-Nov-2013 for paper reference TOPQ-2012-17
%\documentclass[11pt]{article}
%\usepackage{a4wide}\begin{document}
\begin{flushleft}
{\Large The ATLAS Collaboration}

\bigskip

G.~Aad$^{\rm 48}$,
T.~Abajyan$^{\rm 21}$,
B.~Abbott$^{\rm 112}$,
J.~Abdallah$^{\rm 12}$,
S.~Abdel~Khalek$^{\rm 116}$,
O.~Abdinov$^{\rm 11}$,
R.~Aben$^{\rm 106}$,
B.~Abi$^{\rm 113}$,
M.~Abolins$^{\rm 89}$,
O.S.~AbouZeid$^{\rm 159}$
% ...
\bigskip
\\
% ...
$^{11}$ Institute of Physics, Azerbaijan Academy of Sciences, Baku, Azerbaijan\\
$^{12}$ Institut de F{\'\i}sica d'Altes Energies and Departament de F{\'\i}sica de la Universitat Aut{\`o}noma de Barcelona, Barcelona, Spain\\
% ...
$^{21}$ Physikalisches Institut, University of Bonn, Bonn, Germany\\
% ...
$^{48}$ Fakult{\"a}t f{\"u}r Mathematik und Physik, Albert-Ludwigs-Universit{\"a}t, Freiburg, Germany\\
% ...
$^{89}$ Department of Physics and Astronomy, Michigan State University, East Lansing MI, United States of America\\
% ...
$^{106}$ Nikhef National Institute for Subatomic Physics and University of Amsterdam, Amsterdam, Netherlands\\
% ...
$^{112}$ Homer L. Dodge Department of Physics and Astronomy, University of Oklahoma, Norman OK, United States of America\\
$^{113}$ Department of Physics, Oklahoma State University, Stillwater OK, United States of America\\
% ...
$^{116}$ LAL, Universit{\'e} Paris-Sud and CNRS/IN2P3, Orsay, France\\
% ...
$^{159}$ Department of Physics, University of Toronto, Toronto ON, Canada\\
% ...
\end{flushleft}

%%
%% End of file `elsarticle-template-1-num.tex'.
\end{document}